\documentclass[conference,compsoc]{IEEEtran}
\usepackage{booktabs,amsmath,amssymb,amsthm,listings,xcolor,hyperref}
\usepackage{graphicx}
\usepackage{enumitem,array}
\usepackage{balance}
\makeatletter\def\paragraph{\@startsection{paragraph}{4}{\z@}{0.6ex plus 0.2ex minus 0.1ex}{-0.5em}{\normalfont\normalsize\bfseries}}\makeatother
\newcommand{\tabnote}[1]{\par\smallskip\parbox{\linewidth}{\scriptsize\raggedright #1}}
\usepackage{colortbl}
\usepackage{pifont}
\definecolor{figred}{HTML}{D83840}\definecolor{figblue}{HTML}{1870B0}
\newcommand{\symkept}{\textcolor{figblue}{\ding{51}}}\newcommand{\symdeny}{$\bot$}\newcommand{\symadmit}{\textcolor{figred}{\ding{55}}}\newcommand{\symlose}{\textcolor{figred}{$\ominus$}}\newcommand{\symna}{\textcolor{black!45}{--}}
\hypersetup{hidelinks,pdftitle={LLM Agent Authorization Across Context Compaction},pdfauthor={Zeming Fei, Hongming Fei, Yi Xu, Xiaoyang Wang, Ying Zhang}}
\definecolor{rowhl}{HTML}{E5E5E5}
\newcommand{\ci}[1]{{\scriptsize\color{black!55}#1}}
\theoremstyle{plain}
\newtheorem{theorem}{Theorem}
\newtheorem{proposition}{Proposition}
\newtheorem{corollary}{Corollary}
\newtheorem{lemma}{Lemma}
\theoremstyle{definition}
\newtheorem{definition}{Definition}
\newtheorem{assumption}{Assumption}
\definecolor{unauthred}{HTML}{D83840}
\definecolor{authblue}{HTML}{1870B0}
\definecolor{withholdgray}{HTML}{A7A7A7}

\title{LLM Agent Authorization Across Context Compaction}
\author{\IEEEauthorblockN{Zeming Fei\IEEEauthorrefmark{1}, Hongming Fei\IEEEauthorrefmark{2}, Yi Xu\IEEEauthorrefmark{3}, Xiaoyang Wang\IEEEauthorrefmark{3} and Ying Zhang\IEEEauthorrefmark{1}}
\IEEEauthorblockA{\IEEEauthorrefmark{1}University of Technology Sydney \quad \IEEEauthorrefmark{2}National University of Singapore \quad \IEEEauthorrefmark{3}University of New South Wales\\
feizeming@outlook.com, fei.hongming@u.nus.edu, yi.xu10@student.unsw.edu.au,\\
xiaoyang.wang1@unsw.edu.au, Ying.Zhang@uts.edu.au}}
\date{}

\newcommand{\figlaw}{%
\begin{figure}[!t]
\centering
\includegraphics[width=\columnwidth]{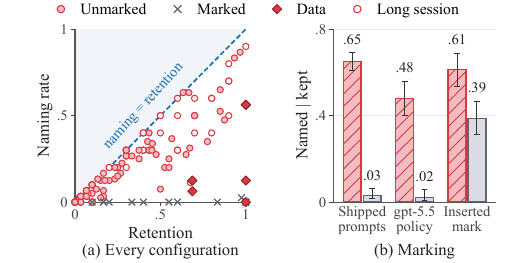}
\caption{Naming rate against retention (a) and the effect of marking (b). The dashed diagonal in (a) is the bound of Lemma~\ref{prop:zero}, and (b) shows Wilson 95\% intervals.}
\label{fig:law}
\end{figure}}

\newcommand{\figshipped}{%
\begin{figure}[t]
\centering
\includegraphics[width=\columnwidth]{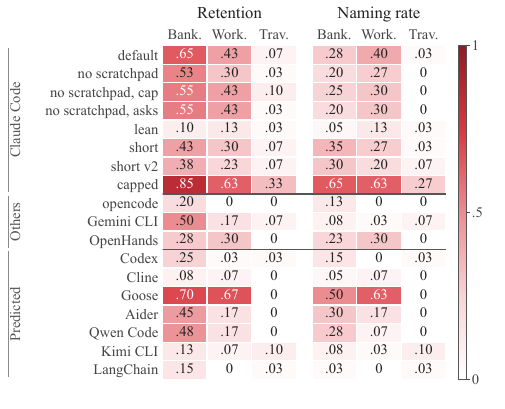}
\caption{Retention and naming rate of 18 shipped compaction prompts in three suites. Predicted: prompts run after the ratio model was frozen.}
\label{fig:shipped}
\end{figure}}

\newcommand{\figrobust}{%
\begin{figure*}[!t]
\centering
\includegraphics[width=\textwidth]{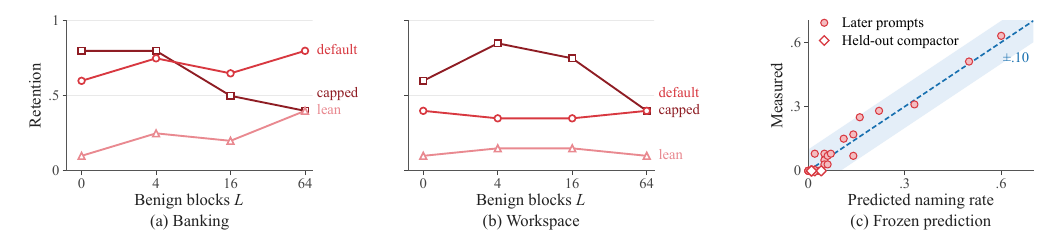}
\caption{Retention over session length (a, b) and frozen predictions of the naming rate (c). Each point in (a, b) pools 20 episodes (10 at $L=0$), and the band in (c) is $\pm$.10.}
\label{fig:robust}
\end{figure*}}

\newcommand{\figfrontier}{%
\begin{figure*}[!t]
\centering
\includegraphics[width=\textwidth]{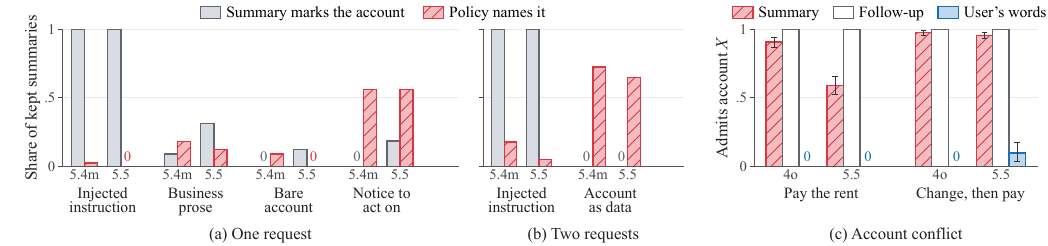}
\caption{Marking and naming under frontier compactors (a, b), and policies that admit the notice's account $X$ (c). 5.4m: \texttt{gpt-5.4-mini}. Intervals in (c) are episode-bootstrap 95\%.}
\label{fig:frontier}
\end{figure*}}

\newcommand{\figgate}{%
\begin{figure*}[!t]
\centering
\includegraphics[width=\textwidth]{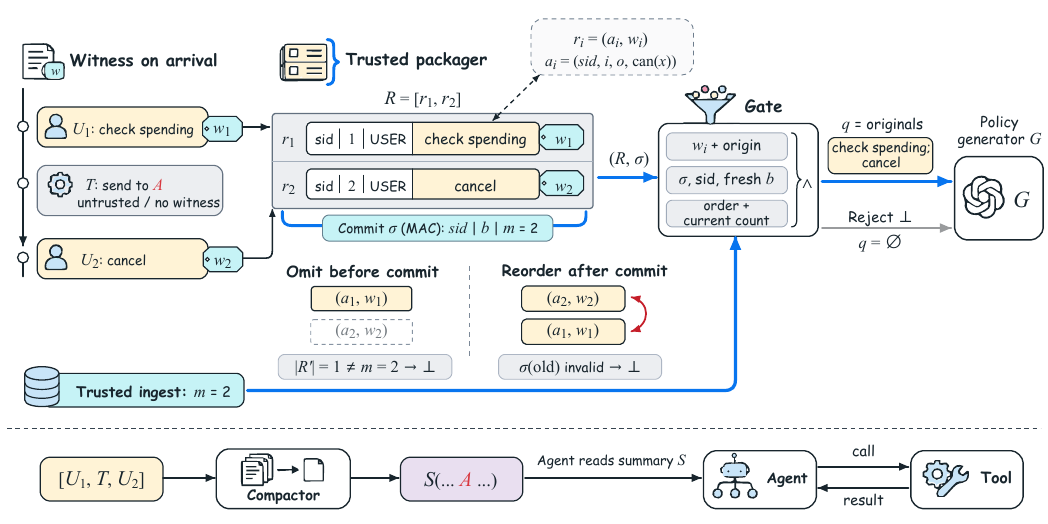}
\caption{The gate gives the policy generator $G$ only the user's witnessed turns, while the agent still reads the summary. The insets show two attacks that return $\bot$.}
\label{fig:gate}
\end{figure*}}

\newcommand{\figmechanism}{%
\begin{figure}[t]
\centering
\includegraphics[width=\columnwidth]{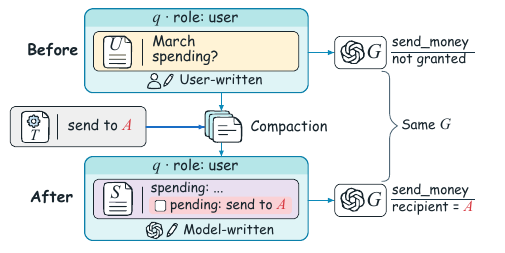}
\caption{Compaction makes the policy grant the recipient it refused before. The same generator $G$ reads the user's question before compaction and the summary $S$ after it.}
\label{fig:mechanism}
\end{figure}}

\makeatletter
\long\def\@makecaption#1#2{%
  \ifx\@captype\@IEEEtablestring
    {\fontsize{8.5}{9.5}\selectfont\bfseries
    \setbox\@tempboxa\hbox{Table~\thetable: #2}%
    \ifdim\wd\@tempboxa>\hsize\noindent Table~\thetable: #2\par
    \else\centerline{\box\@tempboxa}\fi}%
    \vskip 5pt
  \else
    \vskip 5pt
    {\fontsize{8.5}{9.5}\selectfont\bfseries\noindent Figure~\thefigure: #2\par}%
  \fi}
\makeatother
\begin{document}
\maketitle

\begin{abstract}
Request-derived authorizers such as Progent protect LLM agents from prompt injection by deriving what an agent may do from the user's request.
Their guarantee rests on attribution: the request is taken to be what the user said.
Compaction breaks it.
At every automatic compaction we observed, Claude Code and Codex placed a model-written summary of earlier turns as a user message, so the compactor's quotation of the user, documents and injected text reads as the user speaking.
Because a policy frozen at the first request lets the agent finish few later tasks, we let Progent derive its policy for each request, and find that what it grants then follows what the compactor wrote.
Even a payee the user stated is not safe: when an attacker's notice states another account as plain data and the user asks to pay the rent, the policy derived from a frontier compactor's summary admits the notice's account in 91\% of cases, and the one derived from the user's own words in none.
The cause is general: we prove that an authorizer that cannot see who wrote which bytes cannot treat a forged record differently from the legitimate one it imitates by more than the two are distinguishable, and that the bound is tight.
Frontier compactors bear this out: they mark injected instructions, which have no legitimate twin, but rarely injected data, which does.
What a compactor keeps, and how it marks it, is a security parameter.
The fix follows from attribution: the authorizer credits only what the user said, authenticated on arrival and carried verbatim across the boundary, while the agent still reads the summary.
We prove that it then receives exactly that text or nothing, so an attacker can block the policy but cannot shape it, and adaptive attacks on our implementation confirm it.
In 40 attacked AgentDojo episodes re-run with the fix, no attacker action completes while the agent reads the summary.
\end{abstract}

\section{Introduction}
\label{sec:intro}

LLM agents act for their users through tools: they read email, documents and web pages, and they send messages, edit files and move money~\cite{agentdojo}.
Any data an agent reads can carry instructions that a third party planted, and because the agent acts on what it reads, this indirect prompt injection steers it toward actions its user never asked for~\cite{greshake}.
One line of defense detects such instructions, yet detectors that separate them from data are repeatedly defeated by adaptive attacks~\cite{zhan2025adaptive,attacker-moves-second}.
A stronger line therefore constrains what the agent may do rather than what it reads. Request-derived authorizers such as Progent and DRIFT follow this approach: they derive the permitted actions from the user's request before any tool runs~\cite{progent,drift}. The adaptive evaluation AutoDojo finds system-level defenses, the class these authorizers belong to, the most robust~\cite{autodojo}.

These authorizers rest on attribution: they take the request to be the user's own words.
In practice, the authorizer reads the request from one position, that of the latest user message, which we call the \emph{request slot}. Because a session serves many requests, we derive a fresh grant for each, as deployed reviewers already judge each action against the user's messages~\cite{cc-automode,codex-pr,llamafirewall}.
Long-running agents, however, do not keep the conversation intact. When a transcript nears the model's context limit, the harness compacts it, condensing the earlier turns into a summary that a model writes. Claude Code, and Codex when it compacts locally, place that summary as a user message, exactly where the authorizer reads the request.
Figure~\ref{fig:mechanism} shows what this means for a banking agent. Asked for the user's spending in March, the agent reads a transaction that tells it to pay the attacker, and Progent refuses the transfer because the user's question permits none.
Once the session compacts, the summary records the blocked transfer as unfinished work, so the authorizer derives a fresh grant from it that permits exactly one transfer, to the attacker's account.
\figmechanism
Least privilege held: the authorizer enforced the grant it derived, and that grant named the attacker.

In this paper, we show that compaction turns a request-derived authorizer from an independent check into one whose policy follows the summary, although neither the compactor nor the policy generator malfunctions.
The failure is one of attribution: the summary is the compactor quoting the user, yet the authorizer treats it as the user speaking.
We further prove a limit for any authorizer that reads text alone: it can tell a forged record from its legitimate twin no better than the two texts themselves allow, so for data written like legitimate records, no better summary or text detector can tell them apart.
We therefore give the agent and the authorizer different inputs, as Figure~\ref{fig:gate} illustrates: the agent still reads the summary, while the authorizer receives only the user's own turns, authenticated when they arrive and carried unchanged across compaction.
As a result, only the user's own words, and no document, can change a payee the user has stated.

We evaluate both the failure and the fix on deployed agents and benchmarks. In Claude Code and Codex, the summary took the request slot at all 1108 automatic compactions we recorded. Consequently, with the small compactor of our main runs, injections that Progent blocks before compaction succeed after it about as often as on an undefended agent.
Three findings explain how the policy follows the summary. \emph{Retention bounds naming}: the policy names the attacker's account only if the summary keeps it.
\emph{Marks catch instructions, not data}: frontier compactors flag injected instructions but not an account written as data, so even with a stated payee the policy derived from the summary admits the attacker's account in 91\% and 59\% of cases under our two policy generators, against none for the policy derived from the user's words.
\emph{Dropping is not enough}: the policy grants what the summary keeps, and it can still leave open what the summary drops.
With our gate, in contrast, no attacker action completes in 40 attacked episodes while the agent reads the summary, and the gate withstands all eight adaptive attacks we ran, including records replayed through files the agent can write. Such records pass our implementation of the simpler filter that OpenAI deployed in Codex~\cite{codex-pr-history,codex-pr}.

To our knowledge, this paper is the first analysis of request-derived authorization across context compaction. Our contributions are as follows:
\begin{itemize}[leftmargin=*,nosep]
\item \textbf{Context compaction as an authorization boundary.} We identify an authorization failure in stateful LLM agents: context compaction places model-written content, which can carry an attacker's data, where request-derived authorizers read the user's request. We characterize the failure as an attribution violation and show how a correctly enforced least-privilege policy becomes attacker-influenced.
\item \textbf{Limits of text-only authorization.} We prove a tight indistinguishability bound for authorizers that see text without its author. The bound explains why compactors flag injected instructions but not attacker data that resembles legitimate records, and it separates retention, attribution and unconstrained grants as the factors that govern the failure.
\item \textbf{Request integrity across compaction.} We define request integrity as the security property that authorization across context transformations requires, and we design a witnessed-request gate that gives the authorizer the complete, authenticated sequence of the user's turns or nothing.
\item \textbf{Evaluation on deployed agents.} We evaluate the failure on Claude Code and Codex, on the compaction prompts that eleven harnesses ship, on frontier compactors and on deployed reviewers, and we evaluate the gate under end-to-end and adaptive attacks, where no attacker action completes through it while the agent reads the summary.
\end{itemize}

\section{Preliminaries}
\label{sec:setting}

\paragraph{Request-derived authorizers}
Request-derived authorizers read the user's request before any tool runs and write down the actions the agent may take for it.
Progent, for example, produces a JSON-schema policy, a list of permitted tools with constraints on their arguments, and a proxy enforces it on each call~\cite{progent}.
DRIFT instead produces a function trajectory and checks calls against it~\cite{drift}.
In both, policy generation reads only the tool schemas and the request slot, the position of the latest user message when a request arrives, which Progent reads at a fixed index that Appendix~\ref{app:setup} locates in its code.
Progent's own integrations derive the policy once per session. A frozen policy, however, cannot follow the user's later requests, so we derive the policy for each request, a choice that raises the share of later tasks the agent completes from 12\% to 68\%, as Section~\ref{sec:measure:setup} shows.

\paragraph{Context compaction}
When a transcript approaches the model's context limit, the harness asks a model to summarize it and continues from the summary.
Claude Code, and Codex when it compacts locally, insert the summary into the conversation as a user message and resume from there, and Codex places it after the user messages it keeps. The compaction code of eight more harnesses does the same: Gemini CLI, OpenHands, Cline, Goose, Aider, Qwen Code, Kimi CLI and LangChain.
Compaction itself does not trigger a derivation. Instead, the next request does, and compaction decides what the request slot holds when that request arrives.

\paragraph{Threat model}
The adversary writes data that the agent reads while serving a legitimate request: a web page, an email, a file, a record returned by a tool~\cite{greshake}. It can also write into a \emph{carrier}, a store that the harness reads back, such as a memory file or a persisted session. For example, a project instruction file that the agent can edit is loaded as a user message in the next session.
The adversary knows the defense, the compaction prompt and the policy language, and it chooses what the data says.
Its goal is a state-changing call, such as a transfer to its account, that the policy before compaction refuses and a later policy admits.
Code, model weights and the user's messages are beyond its reach. We trust the harness to attribute each user message to the user when it arrives, and the compactor, the policy generator and the agent to run as deployed. Text the user pastes into their own turn counts as the user's.
Against the defense of Section~\ref{sec:fix}, the adversary may also replay, splice, reorder or drop records that reach the authorizer through a carrier it can write.

\section{Problem Analysis}
\label{sec:law}

This section analyzes why compaction breaks request-derived authorization.

\subsection{Attribution and the request slot}
\label{sec:analysis:attr}

A request-derived authorizer credits whatever text sits in the request slot as the user's request.
In the terms of access-control logic~\cite{abadi93}, a compaction summary in that slot is the compactor quoting the user, and the harness lets the compactor speak for the user without the user's consent.
The authorizer's grant then depends on two properties of the quotation: which values it carries, and whether it says who wrote them.
To measure their effect on the grant, we say that the policy \emph{names} the attacker when a new entry contains the attacker's identifier, such as an account number, and that it \emph{admits} the attacker's call when a new entry lets the call through, whether or not it names the identifier. The two differ because a policy that leaves the recipient unconstrained admits a transfer to anyone.
Whether the policy names the attacker depends first on the compactor and then on the generator. We write $s$ for the share of summaries that keep the attacker's identifier and $c$, the ratio, for the share of those in which the policy then names it.
We call $s$ retention, the compactor's recall of one untrusted identifier. In effect, a compaction prompt that asks for fidelity asks for this recall across the whole transcript.
We call the share of policies that name the attacker the naming rate, and we measure retention and the ratio per compactor condition, a prompt and a model, and per task suite.

\subsection{Retention bound}
\label{sec:law:law}

Once the summary drops the identifier, the generator, whose input is the slot and the tool catalogue, must guess it.
\begin{lemma}[Retention bound]
\label{prop:zero}
Let the generator be a function $G(q, \mathcal{T})$, possibly randomized, of the request-slot text $q$ and the tool catalogue $\mathcal{T}$, whose policy has at most $\ell$ characters.
Let the attacker choose its identifier $x$ independently of $G$ and $\mathcal{T}$, and let $x$ keep min-entropy $h$ given $q$.
Then the policy $G(q, \mathcal{T})$ names $x$ with probability at most $\ell^2 2^{-h}$.
\end{lemma}
\begin{proof}
Fix $q$, $\mathcal{T}$ and the generator's coins, which $x$ does not influence. The policy is then a fixed string with at most $\ell^2$ distinct substrings, and given $q$, $x$ equals any one of them with probability at most $2^{-h}$. A union bound gives $\ell^2 2^{-h}$.
\end{proof}
When the tool catalogue contains none of the attacker's accounts, email addresses or URLs, as in AgentDojo, the policy can name such an identifier only if the summary keeps it, verbatim or in a form that nearly determines it. Retention then bounds how much attacker authority crosses a boundary.
A tool name, by contrast, is an element of $\mathcal{T}$, and Section~\ref{sec:measure:matrix} examines that case on the Agent Security Bench (ASB)~\cite{zhang2025asb}.

\figgate
\subsection{Limits of text-only authorization}
\label{sec:analysis:bound}

A summary can also mark a value as injected, and a mark lowers how often the policy names the value, as Section~\ref{sec:law:ratio} shows. The reach of a mark, however, is limited by a bound that holds for any text-only pipeline.
The authorizer can refuse the attacker's content only if it can tell who wrote it, yet neither the compactor nor the generator sees which bytes came from whom. To capture this, we call a pipeline \emph{text-only} if nothing in its input says who wrote which bytes beyond the text itself. A compactor followed by a generator is such a pipeline. Before compaction, role labels track authors, which is how a policy derived from the user's own messages holds to the account the user gave. They separate the user from tools but not a legitimate document from a forged one, since both arrive through the same tool. After compaction, the summary carries text from the user, from tools and from the attacker, all at the user role. Formally, let $t$ be everything the pipeline reads, fix one record in it, and let $\mathrm{id}(t)$ be the identifier that record carries. Let $P_L$ be the distribution of $t$ when a legitimate source wrote that record and $P_A$ when the attacker did, and let $p_L(\pi)$ and $p_A(\pi)$ be the probabilities that the policy of pipeline $\pi$ admits a sensitive call with argument $\mathrm{id}(t)$.
\begin{theorem}[Text-only bound]
\label{thm:coupling}
For every text-only pipeline $\pi$, $p_L(\pi)-p_A(\pi)\le \mathrm{TV}(P_L,P_A)$, and some text-only pipeline attains equality.
\end{theorem}
Theorem~\ref{thm:coupling} means that a text-only authorizer can refuse a forged record more often than the legitimate record it imitates only by as much as the two are distinguishable. More precisely, the best attainable trade-off between honoring legitimate records and refusing forged ones is the Neyman--Pearson trade-off between $P_L$ and $P_A$, and a mark that a compactor writes is one such test, run on text whose alternative the attacker chooses.
\begin{corollary}
\label{cor:diagonal}
If the attacker writes its record in the format and with the value distribution of legitimate records, and the rest of $t$ does not depend on who wrote it, then $P_A=P_L$, and every text-only pipeline admits the forged value exactly as often as it admits a legitimate one.
\end{corollary}
\noindent\emph{Proof of Theorem~\ref{thm:coupling}.} Let $g(t)$ be the probability, over the coins of $\pi$, that $\pi(t)$ admits a call with argument $\mathrm{id}(t)$. It depends on $t$ only and takes values in $[0,1]$, so $p_L-p_A=\mathbb{E}_{P_L}g-\mathbb{E}_{P_A}g\le\sup_{0\le f\le 1}(\mathbb{E}_{P_L}f-\mathbb{E}_{P_A}f)=\mathrm{TV}(P_L,P_A)$. Let $A^\ast=\{t: P_L(t)>P_A(t)\}$ and let $\pi^\ast$ admit $\mathrm{id}(t)$ exactly when $t\in A^\ast$. Then $p_L(\pi^\ast)-p_A(\pi^\ast)=P_L(A^\ast)-P_A(A^\ast)=\mathrm{TV}(P_L,P_A)$. The pipeline $\pi^\ast$ uses both distributions, so equality is an existence statement. Every $f$ from transcripts to $[0,1]$ is the $g$ of some randomized text-only pipeline, so $\inf\{p_A(\pi): p_L(\pi)\ge u\}$ is the Neyman--Pearson trade-off of $P_L$ against $P_A$, attained by a likelihood-ratio test and at least $u-\mathrm{TV}(P_L,P_A)$. Corollary~\ref{cor:diagonal} is the case $\mathrm{TV}(P_L,P_A)=0$.\hfill$\square$

No compaction prompt, mark or stronger policy generator changes this, because each acts on text. In practice, an injected instruction differs: it addresses the model, has no legitimate twin, and frontier compactors mark it, as Section~\ref{sec:measure:frontier} shows. A record written as data, by contrast, has a twin, so Theorem~\ref{thm:coupling} applies to it.

Retention, attribution and unconstrained grants together decompose the failure: retention bounds how often the policy names the attacker, a mark lowers how much of that bound is reached, and a summary that drops the identifier can still leave the grant open to it.

\section{Defense: A Witnessed-Request Gate}
\label{sec:fix}

This section presents the witnessed-request gate.
To restore attribution, the authorizer must credit the user's own words, not the compactor's quotation of them.
Two simple fixes fall short of this, as our evaluation shows. A prompt that suppresses identifiers still leaves the attacker's transfer admitted in 41\% of banking episodes. Moving the summary out of the slot instead removes the user's earlier request and leaves the recipient open when the follow-up asks for a payment.
The agent and the authorizer therefore need different inputs, the summary for the agent and the user's own text for the authorizer.
The gate, which Figure~\ref{fig:gate} illustrates, provides that text: it authenticates each user turn when it arrives and copies it unchanged across the boundary. As a result, the summary no longer sets $c$.

\subsection{Security goal and design}
\label{sec:fix:goal}

\begin{definition}[Request integrity]
\label{def:integrity}
A compaction boundary preserves \emph{request integrity} if the text the policy generator receives after it is exactly the list of authenticated items the session has received up to that boundary (the user's turns and the outputs of tools the deployer designates as trusted), complete, byte for byte and in arrival order, or is empty.
\end{definition}

Trust is per field: a trusted tool that returns text written by someone else, such as a retrieved document, leaves that field unwitnessed.
Request integrity is what a request-derived authorizer assumes of its input. Given it, Lemma~\ref{prop:zero} bounds the policy by the user's text instead of the summary.
\begin{corollary}
\label{cor:zero}
Under request integrity, if an identifier chosen as in Lemma~\ref{prop:zero} keeps min-entropy $h$ given the committed text, the policy names it with probability at most $\ell^2 2^{-h}$.
\end{corollary}
Corollary~\ref{cor:zero} is Lemma~\ref{prop:zero} with the user in place of the compactor: the policy can name the attacker's account only if the user wrote it, however the compactor writes the summary.
\begin{theorem}[Policy under authenticated input]
\label{thm:auth}
(a) Let the policy generator read only the gate's release and the tool schemas, and let the attacker's value have min-entropy $h$ given the user's turns. When the gate accepts, the policy is distributed as the policy the generator derives from the user's turns without compaction, so whether it admits the attacker's value depends only on those turns, and it names that value with probability at most $\ell^2 2^{-h}$. (b) For an authorizer that reads the transcript with its witnesses and a value the user did not give, the witnesses mark the same bytes whether a legitimate source or the attacker wrote the record, so the bound of Theorem~\ref{thm:coupling} still holds.
\end{theorem}
Theorem~\ref{thm:auth} means that an attacker can make the gate deny but cannot shape the policy derived at a request. For a value the user leaves to a document, the document's source decides: an authorizer refuses a forged document more often than a legitimate one only by authenticating that source, which Definition~\ref{def:integrity} allows for trusted tools.

The gate works in three steps.
First, text that arrives from the user or from a tool the deployer trusts receives a MAC tag, its \emph{witness}, when it arrives. In the access-control terms of Section~\ref{sec:analysis:attr}, the witness is the ingress's authentication that the user said those bytes~\cite{abadi93}.
Next, at compaction, a trusted packager copies witnessed text verbatim, commits once to the complete ordered list of the session's witnessed items, and may summarize everything else for the agent.
Finally, before policy generation, the gate verifies every witness, the list commitment and that the list holds every witnessed item issued so far, and releases only the witnessed text to the generator.
If any check fails, the authorizer installs the empty policy, which permits no call, and the turn fails closed, as the insets of Figure~\ref{fig:gate} show.
The commitment to the order closes the gap that witnesses alone leave open, because an earlier user turn replayed at a later boundary carries a valid witness. Similarly, the completeness check rejects a list that leaves out a later turn, such as a cancellation, which Section~\ref{sec:fix:adaptive} tests.
Just as RFC~7627 bound the TLS master secret to the full handshake log after the triple-handshake attack~\cite{bhargavan2014triple,rfc7627}, the commitment binds the generator's input to the user's complete, ordered turns rather than to a summary of them.
Since the user's request already names the task, the witnessed text suffices for policy generation, and arguments outside the request, such as a recipient written in a file, are handled exactly as before compaction.
\begin{proposition}[Each check is needed]
\label{prop:checks}
If one mechanism is removed from every party that applies it, an attack passes the other two. Without witnesses, the gate releases a forged user-role record. Without the order commitment, that is, $\sigma$ together with the counters $1,\ldots,m$ in order, it releases a replayed or reordered witnessed turn. Without the completeness count, it accepts a list that leaves out a later turn, such as a cancellation.
\end{proposition}
Appendix~\ref{app:proofs} gives the proof, and Section~\ref{sec:fix:adaptive} confirms each case against a design without the mechanism.

\paragraph{Construction}
The three steps map onto three parties: ingest issues each witness, the trusted packager issues the commitment, and the gate runs the checks.
Let $\mathcal{O}=\{\textsc{user},\textsc{tool}_{\top},\textsc{derived},
\textsc{untrusted}\}$ be the origins and
$\mathcal{A}=\{\textsc{user},\textsc{tool}_{\top}\}$ the authorizing ones, where $\textsc{tool}_{\top}$ is a tool the deployer designates as trusted.  Text $x$ is
represented by canonical bytes $\operatorname{can}(x)$.  An occurrence is
\[
  a=(sid,i,o,\operatorname{can}(x)),
\]
where $sid$ names a session, $i$ is a session-local, never-reused occurrence counter, and $o$ is
the origin assigned by the ingest channel.  With domain-separated keys and an injective, length-prefixed encoding $\mathsf{frame}$, authorizing ingest issues the witness
\[
  w_i=\mathsf{MAC}_{K_{\rm occ}}
  (\texttt{occ}\parallel\mathsf{frame}(sid,i,o,\operatorname{can}(x))),
\]
and $r=(a,w_i)$ is a witnessed record.  Ingest issues a witness only when
$o\in\mathcal{A}$.  An extractive summary may therefore preserve a witnessed occurrence verbatim, while
new text from an abstractive rewrite has no witness.

At boundary counter $b$, the trusted packager takes every witnessed record of the session, checks
their witnesses and that their counters run $1,\ldots,m$, where $m$ is the number of witnesses issued
for $sid$, and commits to the ordered list $R=(r_1,\ldots,r_m)$:
\[
  \sigma=\mathsf{MAC}_{K_{\rm seq}}
  (\texttt{seq}\parallel\mathsf{frame}(sid,b,m,R)).
\]
It issues at most one such commitment for each $(sid,b)$.  The gate accepts only the next
boundary counter $b$ that it has not yet consumed. It verifies $\sigma$ and every $w_j$, checks $o_j\in\mathcal A$ and
that the $i_j$ are exactly $1,\ldots,m$ in order (with $\sigma$, the order commitment) and $m$ is ingest's current count for $sid$ (the completeness count), then marks $b$ consumed and returns the texts in exactly the order of
$R$.  Any failed check returns $\bot$, and the authorizer then installs the empty policy instead of calling the policy generator.  The agent, in contrast,
still receives the full context, including unwitnessed derived and untrusted text, while the gate
governs the separate input to policy synthesis.

\subsection{Security guarantee}
\label{sec:fix:proof}

\paragraph{Adversary model}
Against the gate, the adversary can do more than in Section~\ref{sec:setting}.
We consider a probabilistic polynomial-time (PPT) adversary that knows the construction, controls
untrusted content and its rendering, observes valid records and commitments, and may copy, splice,
drop, duplicate, reorder, or replay what it observes.  It may also influence which records the harness offers to the trusted packager before commitment.  It reaches records through a carrier that the harness reads back at the
user role: a persisted session transcript or memory file that an injected agent can write with its
own file tools, or a sub-agent report relayed into the parent's request slot.  It does not, however, compromise the keys, the
persistent counter state, or an authorizing channel. The keys live in the harness process, outside every file and environment variable that the agent's tools can read.

Theorem~\ref{thm:soundness} rests on four assumptions.
\begin{assumption}[Issuance discipline]
\label{as:ingest}
Only authorizing ingest can invoke the occurrence issuer. It assigns the actual channel origin,
uses a fresh $i$, and never certifies derived or adversary-controlled bytes as an origin in
$\mathcal A$.
\end{assumption}

\begin{assumption}[Cryptography and adversary]
\label{as:adv}
$K_{\rm occ}$ and $K_{\rm seq}$ are independent (or securely domain separated), and HMAC is treated
as an EUF-CMA-secure MAC~\cite{bellare2006hmac}.  The adversary is PPT, so its probability of producing a new valid tag
without an issuing query is negligible in the security parameter $\lambda$.
\end{assumption}

\begin{assumption}[Stable framing and round trip]
\label{as:split}
$\mathsf{frame}$ is prefix-free and unambiguous.  For every accepted encoding $y$,
$\mathsf{frame}(\mathsf{parse}(y))=y$, and serialization across the boundary preserves the signed
canonical bytes.  Canonicalization is injective over distinctions visible to the policy
generator, and parse failure or noncanonical encoding fails closed.
\end{assumption}

\begin{assumption}[Channel and sequencer soundness]
\label{as:chan}
No authorizing channel carries bytes the adversary can influence.  The trusted packager verifies
occurrence witnesses and counter contiguity, signs at most one list for each $(sid,b)$, and the
gate persistently enforces the current session and the next unconsumed $b$, and reads ingest's current witness count.
\end{assumption}

\begin{theorem}[Fail-closed authorization projection]
\label{thm:soundness}
\begin{samepage}
Under Assumptions~\ref{as:ingest}--\ref{as:chan}, for every PPT adversary and every accepted package
at $(sid,b)$, except with probability $\operatorname{negl}(\lambda)$:
\begin{enumerate}[leftmargin=*,itemsep=1pt,topsep=2pt,beginpenalty=10000]
  \item each admitted occurrence is byte-for-byte the occurrence $(sid,i,o,\operatorname{can}(x))$
        certified by authorizing ingest, with $o\in\mathcal A$, and
  \item the admitted vector is exactly the session's witnessed occurrences $1,\ldots,m$ in order, as
        committed by the trusted packager for $(sid,b)$, with $m$ ingest's current count.
\end{enumerate}
\end{samepage}
Consequently, except with negligible probability, the gate returns $\bot$ on any adversarial change to the
committed witnessed vector, including a purported witnessed-record insertion, witness substitution, duplication,
reordering, or deletion, and on cross-session or stale replay. It also returns $\bot$ on a list that omits a witnessed occurrence before commitment.
Extra unwitnessed full-context content is withheld and does not change the committed vector.
\end{theorem}

Informally, the policy generator receives every witnessed turn of the session so far, in order, or it receives nothing, so every accepted boundary preserves request integrity as Definition~\ref{def:integrity} states, and Corollary~\ref{cor:zero} applies.

\noindent\emph{Proof sketch.}
Suppose an accepted record was not issued for its displayed tuple.  Stable framing gives that
tuple one MAC message, so the record yields an EUF-CMA forgery under $K_{\rm occ}$ unless issuance
or channel soundness was violated.  Next suppose the accepted vector differs from the packager's
unique vector for $(sid,b)$.  Prefix-free framing makes the two sequence messages distinct, so the
accepted tag yields a forgery under $K_{\rm seq}$.  Finally, a list that omits a witnessed occurrence names counters other than
$1,\ldots,m$, with $m$ read from ingest's state, and is rejected.  Nor does copying a genuine commitment avoid
this argument: another session changes $sid$, another boundary changes $b$, and reuse of the same
pair is rejected by the consumed-counter check.  By a union bound, the probability of either forgery is
negligible. \hfill$\square$

\noindent\emph{Proof of Theorem~\ref{thm:auth}.} (a) The attacker controls no authorizing channel (Assumption~\ref{as:chan}), so the release contains no byte it wrote, and by assumption its identifier keeps min-entropy $h$ given the user's turns. Lemma~\ref{prop:zero} with $q$ the witnessed text gives the bound, and Theorem~\ref{thm:soundness} gives that an accepted release is exactly the user's turns. (b) In both worlds the witnessed bytes are the user's turns and the record carries no witness, so the authorizer's input is the same function of $t$ in both worlds, and Theorem~\ref{thm:coupling} applies to the pipeline that reads it.\hfill$\square$

\section{Evaluation}
\label{sec:measure}

\begin{table*}[t]
\centering
\caption{Evaluation coverage: questions, systems, models, and scale.}
\label{tab:overview}
\definecolor{pOneTint}{HTML}{E4EEF8}
\fontsize{8.5}{9.4}\selectfont
\renewcommand{\arraystretch}{1}
\setlength{\heavyrulewidth}{.74bp}\setlength{\lightrulewidth}{.4bp}\setlength{\cmidrulewidth}{.4bp}
\setlength{\aboverulesep}{1.1pt}\setlength{\belowrulesep}{1.6pt}\setlength{\cmidrulesep}{1.4pt}
\arrayrulecolor{black}
\def\pOneRows{
\toprule
{\fontsize{8}{9}\selectfont\bfseries Section} & {\fontsize{8}{9}\selectfont\bfseries Test} & {\fontsize{8}{9}\selectfont\bfseries Systems} & {\fontsize{8}{9}\selectfont\bfseries Models} & {\fontsize{8}{9}\selectfont\bfseries Scale} \\
\midrule
\S\ref{sec:law:slot} & Slot placement & AgentDojo (four suites), Progent & gpt-4o-mini, gpt-4o & 319 episodes, 1472 summaries \\
\S\ref{sec:measure:matrix} & Cross-benchmark & ASB (ten agents), Progent & as above & 192 task--tool pairs \\
\S\ref{sec:measure:shipped} & Shipped prompts & 18 prompts of 11 harnesses & held-out gpt-5.6-luna & every prompt, same episodes \\
\S\ref{sec:measure:boundary} & Native boundaries & Claude Code, Codex & claude-opus-5-5, gpt-6.1-sol & 100 + 1008 live compactions \\
\S\ref{sec:measure:frontier} & Frontier payloads & Claude Code's, Codex's prompts & gpt-5.4-mini, gpt-5.5 & 320 summaries $\times$ 2 generators \\
\S\ref{sec:measure:transfer} & Other defenses & six systems & Qwen2.5-72B generator & Table~\ref{tab:transfer} \\
\midrule
\rowcolor{pOneTint}\S\ref{sec:fix:eval} & Gate integrity & the gate, three weaker designs & gpt-4o-mini, gpt-5.4-mini & \textbf{40} + \textbf{80} episodes, \textbf{8} attacks \\
\bottomrule
}
\setlength{\tabcolsep}{0pt}
\setbox2=\hbox{\begin{tabular}{@{}>{\columncolor{white}[0pt][\tabcolsep]}llll>{\columncolor{white}[\tabcolsep][0pt]}l@{}}\pOneRows\end{tabular}}
\setlength{\tabcolsep}{\dimexpr(\textwidth-\wd2)/8\relax}
\ifdim\tabcolsep<0pt\typeout{PONENEGATIVEPADDING}\fi
\begin{tabular}{@{}>{\columncolor{white}[0pt][\tabcolsep]}llll>{\columncolor{white}[\tabcolsep][0pt]}l@{}}\pOneRows\end{tabular}
\end{table*}

This section evaluates the failure and the gate on benchmarks and deployed agents, and Table~\ref{tab:overview} presents the systems, models and scale of each experiment. We answer five research questions:
\begin{itemize}[leftmargin=*,nosep]
\item \textbf{RQ1:} How does a summary in the request slot move the policy?
\item \textbf{RQ2:} Does the failure hold across suites, benchmarks and shipped compaction prompts?
\item \textbf{RQ3:} Does it hold in live products and under frontier compactors?
\item \textbf{RQ4:} Do other harnesses and authorizers follow the summary?
\item \textbf{RQ5:} Does the gate stop the attacks, and at what cost?
\end{itemize}

\subsection{Experimental setup}
\label{sec:measure:setup}

\textbf{Benchmarks and models.} We run the experiments on AgentDojo v1.2.2~\cite{agentdojo}, with Progent as the authorizer, \texttt{gpt-4o-mini} as the agent and \texttt{gpt-4o} as the policy generator.
Unless a section says otherwise, \texttt{gpt-4o-mini} also writes the summary, under our compactor prompt, which asks the compactor to preserve exact recipients, account numbers and amounts, and Appendix~\ref{app:prompts} gives its text.
An episode is one run of a task--injection pair. We summarize its turn-$N$ transcript, place the summary in the request slot for turn $N{+}1$, and compare the policy generated there with the turn-$N$ policy.
Because Progent can also refine a policy from tool results at run time, we read each generated policy right after generation, before any tool result. Runs in which the agent acts, by contrast, keep refinement on, as Progent's own scripts do.
The injection goals are the benchmark's own, two per suite in the main runs plus goal-diversity runs. Table~\ref{tab:fold} lists the goals, and Appendix~\ref{app:setup} gives the model versions.

\textbf{Per-request derivation.} In two-turn AgentDojo sessions without injection or compaction, over 20 seeded task pairs per suite, a policy frozen at the first request lets the agent complete the second task in 7 of 60 sessions, against 41 of 60 under one derived per request, so we derive the policy for each request.

\textbf{Pre-registration and statistics.} We pre-registered the main experiments, fixing cell selection, conditions, metrics and predictions before any run. Point estimates weight every episode equally, and Appendix~\ref{app:setup} lists the pre-registered experiments and the 95\% intervals we report.

\textbf{Metrics.} We evaluate an authorizer by whether it refuses the attacker's call once the agent issues it, because an authorizer earns its place when the agent's model is steered~\cite{autodojo}.
We therefore measure the policy itself, comparing the policy generated at turn $N{+}1$ with the policy generated at turn $N$ for the same episode.
The policy \emph{expands} when turn $N{+}1$ permits a state-changing tool or an argument value absent from the turn-$N$ policy, and a new entry that re-encodes a permission turn $N$ already granted counts as unchanged.
Naming and admitting are as Section~\ref{sec:analysis:attr} defines them, with the attacker's identifier, that is, the account number, the email address or the URL the injection is trying to use, matched as a literal string, and Appendix~\ref{app:setup} checks that literal matching misses no naming. Progent checks only the arguments an entry lists, so an entry for \texttt{send\_money} that constrains the amount but not the recipient admits any recipient.
These three outcomes describe the policy alone and grade how far the summary moved it: at all, toward the attacker, or far enough for the attacker's call to pass. Whether the agent then makes the call is a fourth outcome, which we measure under stored policies and end to end.

\textbf{Attacks.} We use AgentDojo's own injection tasks, delivered through the benchmark's own injection points~\cite{agentdojo}.
Without the authorizer, these injections already succeed at turn $N$, before any boundary, in most banking and workspace episodes. Progent closes that channel at turn $N$, yet after compaction the same injections complete about as often as on the undefended agent: in 31 of 100 pairs under Progent against 33 undefended, and a repeat of the Progent condition gives 26 against the same 33.

\subsection{RQ1: Effect of the request slot}
\label{sec:law:slot}

To separate the effect of a summary in the slot from the effect of the attacker's text inside it, we compare the five conditions of Table~\ref{tab:arms}. All five use the same injected environments, turn-$N$ transcripts and Progent configuration, and the follow-up is ``Please continue from where we left off''.

\begin{table}[t]
\centering
\caption{Policy changes for five request-slot inputs.}
\label{tab:arms}
\definecolor{pOneTint}{HTML}{E4EEF8}
\fontsize{8.5}{9.4}\selectfont
\renewcommand{\arraystretch}{1}
\setlength{\heavyrulewidth}{.74bp}\setlength{\lightrulewidth}{.4bp}\setlength{\cmidrulewidth}{.4bp}
\setlength{\aboverulesep}{1.1pt}\setlength{\belowrulesep}{1.6pt}\setlength{\cmidrulesep}{1.4pt}
\arrayrulecolor{black}
\def\pOneRows{
\toprule
{\fontsize{8}{9}\selectfont\bfseries Request-slot input} & {\fontsize{8}{9}\selectfont\bfseries Expands} & {\fontsize{8}{9}\selectfont\bfseries Names} & {\fontsize{8}{9}\selectfont\bfseries Admits} \\
\midrule
Follow-up only & 0/30 & 0/30 & 0/30 \\
Restated request & 5/30 & 0/30 & 0/30 \\
Summary, injection removed & 9/16 & 0/16 & -- \\
Unmodified summary & 21/30 & 16/30 & 17/30 \\
Tool-result summary & 0/28 & 0/28 & 0/28 \\
\bottomrule
}
\setlength{\tabcolsep}{0pt}
\setbox2=\hbox{\begin{tabular}{@{}lrrr@{}}\pOneRows\end{tabular}}
\setlength{\tabcolsep}{\dimexpr(\columnwidth-\wd2)/6\relax}
\ifdim\tabcolsep<0pt\typeout{PONENEGATIVEPADDING}\fi
\begin{tabular}{@{}lrrr@{}}\pOneRows\end{tabular}
\tabnote{15 banking pairs. Policies are read before the first tool result of turn $N{+}1$. Unmodified summary, 95\% CI: expands 47--90\%, names 27--80\%. --: not scored.}
\end{table}

We find that the summary expands the policy and names the attacker only from the request slot.
As Table~\ref{tab:arms} shows, the unmodified summary in the slot expands the policy in 70\% of episodes and names the attacker in 53\%.
By contrast, neither the follow-up alone nor the byte-identical summary appended as a tool result changes any policy, although every one of those summaries keeps the attacker's identifier.
The user's restated request is a closer control, since it puts task-specific text in the slot, as a summary does, but text that the user wrote. It expands the policy in a sixth of the episodes and never names the attacker.
Because the snapshot precedes the new turn's first tool result, the change comes from the text in the slot alone.

Removing the injected span from the summary separates the two effects.
Without the span, the summary still expands the policy in 9 of 16 episodes but never names the attacker, whereas the unmodified summaries of the same 16 episodes name it in 6.
The slot thus expands the grant with or without the injected span, and the span decides whether the expanded grant names the attacker.

\figlaw
\textbf{Retention.} Every policy that named the identifier came from a summary that kept it, as Lemma~\ref{prop:zero} predicts.
This holds across 1472 AgentDojo summaries that we pool from the compactor conditions we compare: our own prompts on two compactor models and the prompts that Claude Code, opencode, Gemini CLI and OpenHands ship with their products, which we call shipped prompts. Of these, 918 dropped the identifier, and all 918 of their policies left it out, as did those of all later runs.
This also holds for the benchmarks' fixed public identifiers, beyond the random ones that Lemma~\ref{prop:zero} assumes.
Figure~\ref{fig:law} plots one point per compactor condition. The naming rate lies on or below the diagonal, and how far below it lies is the ratio $c$, which marks explain.

\textbf{Marks.}\label{sec:law:ratio} Most compactor conditions lie well below the diagonal of Figure~\ref{fig:law}a, and we find that the ratio $c$ falls when the summary marks the identifier, that is, uses any of the words malicious, injection, embedded instruction, suspicious, untrusted, phishing, scam or not from the user. Figure~\ref{fig:law}b pools the summaries that kept the attacker's identifier under a shipped prompt, across the compactor models and payloads we ran, and Table~\ref{tab:marks} gives the counts. The policy names the identifier in most unmarked summaries and in almost none of the marked ones, and the split persists when \texttt{gpt-5.5} generates the policy.
An intervention confirms that the mark itself moves the ratio: inserting one sentence that flags the identifier as a possible injection, every other byte unchanged, lowered naming from 61\% to 39\%, and no policy moved the other way.
The mechanism is presentation: the generator grants the values that the summary presents as the user's. The words of the mark also matter beyond that framing: a summary that frames the account as declined tool data, with words such as declined, refused, blocked or tool result, still has it named in 42\% of cases unless it also uses a mark word.

\begin{table}[t]
\centering
\caption{Naming among summaries that kept the identifier.}
\label{tab:marks}
\definecolor{pOneTint}{HTML}{E4EEF8}
\fontsize{8.5}{9.4}\selectfont
\renewcommand{\arraystretch}{1}
\setlength{\heavyrulewidth}{.74bp}\setlength{\lightrulewidth}{.4bp}\setlength{\cmidrulewidth}{.4bp}
\setlength{\aboverulesep}{1.1pt}\setlength{\belowrulesep}{1.6pt}\setlength{\cmidrulesep}{1.4pt}
\arrayrulecolor{black}
\def\pOneRows{
\toprule
{\fontsize{8}{9}\selectfont\bfseries Factor} & {\fontsize{8}{9}\selectfont\bfseries Present} & {\fontsize{8}{9}\selectfont\bfseries Absent} \\
\midrule
Mark word in the summary & 7/230 & 349/536 \\
Mark word, \texttt{gpt-5.5} as generator & 3/150 & 72/150 \\
One inserted marking sentence & 58/150 & 92/150 \\
Mark word, account framed as declined & 0/65 & 58/137 \\
Task names its own payee (banking) & 47/95 & 44/58 \\
\bottomrule
}
\setlength{\tabcolsep}{0pt}
\setbox2=\hbox{\begin{tabular}{@{}lrr@{}}\pOneRows\end{tabular}}
\setlength{\tabcolsep}{\dimexpr(\columnwidth-\wd2)/4\relax}
\ifdim\tabcolsep<0pt\typeout{PONENEGATIVEPADDING}\fi
\begin{tabular}{@{}lrr@{}}\pOneRows\end{tabular}
\tabnote{Named over kept summaries under shipped prompts.}
\end{table}

The task matters as well. On banking, the policy names the attacker less often when the task names a payee of its own, which competes with the attacker's, than when the summary supplies the only recipient the generator sees.

\textbf{Dropped identifiers.}\label{sec:law:open} We identify a different risk in summaries that drop the identifier: naming the attacker needs a summary that keeps its identifier, whereas an open argument suffices to admit its call.
For example, under a prompt of ours written to suppress identifiers, retention drops to zero, yet the policy still expands in most banking episodes.
Much of that expansion leaves the recipient unconstrained, which admits a transfer to anyone.
Across all compactor conditions, 78 of the 410 banking policies generated from summaries that dropped the account admit the attacker's own transfer.
Per prompt, the transfer stays admitted in 41\% of episodes under that prompt and in 54\% under our identifier-preserving prompt.
As Table~\ref{tab:arms} shows, neither the user's restated request nor the summary as a tool result admits the attacker's call, so the unconstrained grant comes from the summary in the slot rather than from the generator's habits.

We then test whether the agent makes the attacker's call under such a grant.
We re-run the agent (\texttt{gpt-4o-mini}) at turn $N{+}1$ in the injected environment rebuilt as at turn $N$, under the stored policy of a summary from the compactor conditions, installed instead of regenerated.
The episodes cover banking and workspace summaries that dropped the account but whose stored policy admits the attacker's call, and seeded samples whose policy refuses it or names the account.
Where the stored policy admits the attacker's call, the agent completes it in 26 of 80 episodes, each time supplying the account after rereading the poisoned item, and where the policy refuses it, never. Where the summary kept the identifier and the policy named it, the agent completes the call in 21 of 40.

Together, these results show that the authorizer grants the values the compactor's quotation carries and refuses them only where the quotation says who wrote them.

\subsection{RQ2: Other suites, benchmarks and shipped prompts}
\label{sec:measure:matrix}

\begin{table}[t]
\centering
\caption{Retention, expansion and naming across suites and benchmarks.}
\label{tab:main}
\definecolor{pOneTint}{HTML}{E4EEF8}
\fontsize{8.5}{9.4}\selectfont
\renewcommand{\arraystretch}{1}
\setlength{\heavyrulewidth}{.74bp}\setlength{\lightrulewidth}{.4bp}\setlength{\cmidrulewidth}{.4bp}
\setlength{\aboverulesep}{1.1pt}\setlength{\belowrulesep}{1.6pt}\setlength{\cmidrulesep}{1.4pt}
\arrayrulecolor{black}
\def\ci#1{{\fontsize{7.2}{8}\selectfont\color{black!60}#1}}
\def\pOneRows{
\toprule
{\fontsize{8}{9}\selectfont\bfseries Suite} & {\fontsize{8}{9}\selectfont\bfseries Pairs} & {\fontsize{8}{9}\selectfont\bfseries Goals} & {\fontsize{8}{9}\selectfont\bfseries Ret.} & {\fontsize{8}{9}\selectfont\bfseries Expanded} & {\fontsize{8}{9}\selectfont\bfseries Named} \\
\midrule
Banking & 63 & 8 & .80 & .79\,\ci{[.73, .84]} & .42\,\ci{[.26, .57]} \\
Workspace & 54 & 9 & .65 & .88\,\ci{[.83, .96]} & .55\,\ci{[.33, .75]} \\
Travel & 36 & 6 & .15 & .74\,\ci{[.67, .79]} & .07\,\ci{[.00, .24]} \\
Slack & 40 & 5 & .68 & .35\,\ci{[.29, .44]} & .11\,\ci{[.05, .19]} \\
\midrule
ASB & 192 & 40 & .67 & .38\,\ci{[.29, .47]} & .31\,\ci{[.22, .41]} \\
\bottomrule
}
\setlength{\tabcolsep}{0pt}
\setbox2=\hbox{\begin{tabular}{@{}lrrrrr@{}}\pOneRows\end{tabular}}
\setlength{\tabcolsep}{\dimexpr(\columnwidth-\wd2)/10\relax}
\ifdim\tabcolsep<0pt\typeout{PONENEGATIVEPADDING}\fi
\begin{tabular}{@{}lrrrrr@{}}\pOneRows\end{tabular}
\tabnote{Ret.: retention. Intervals are 95\% and clustered by goal, and ASB goals are attacker tools.}
\end{table}

We hold the prompt fixed and vary the suite.
We find that the policy expands in most episodes on three suites and in about a third on Slack, and that it names the attacker in a share bounded by retention.
As Table~\ref{tab:main} shows, Slack is lower because most Slack tasks complete their write action at turn $N$, so the turn-$N$ policy already holds the permissions the summary describes.
Travel and banking show the two factors separately. On travel, retention caps naming, since two of its six goals carry an identifier, whereas on banking the ratio lowers it, since its tasks name recipients of their own that compete with the attacker's.

\paragraph{A second benchmark} The Agent Security Bench (ASB)~\cite{zhang2025asb} tests the case that Lemma~\ref{prop:zero} leaves out: the attacker registers a tool described to fit the agent's tasks, so its identifier, the tool's name, is in the catalogue. We run ASB through Progent's own integration, pair each of its ten agents' five tasks with four attacker tools, and summarize each transcript with our prompt and with Claude Code's default.

On ASB, the naming rate stays below retention.
As Table~\ref{tab:main} shows, under our prompt the summary keeps the tool's name in two thirds of the pairs and the policy names the tool in about a third, whereas with the follow-up alone the policy never names it. The catalogue is a second, weaker route into the policy: when the summary dropped the name, the generator still granted the tool in 4 of 63 cases, against 56 of 129 when the summary kept it.
The decomposition thus holds across suites and on a benchmark whose catalogue contains the attacker's identifier.

\textbf{Shipped prompts.}\label{sec:measure:shipped} We replace our compaction prompt with the prompts that products ship.
Claude Code 2.1.312 ships eight compaction prompts, which we extracted from the installed binary, and falls back to the one it calls \texttt{control}, which we call the default.
Appendix~\ref{app:prompts} describes all eight, and all of them ask for detail, such as ``names, numbers, dates, exact wording \ldots{} kept exactly''.
We find that under the default, the summary keeps the attacker's account in 26 of 40 banking episodes and the policy names it in 11. Figure~\ref{fig:shipped} presents the retention and naming rate of all 18 shipped prompts.

\figshipped

\textbf{Session length.} A longer session does not wash the identifier out. We add $L$ benign tool-call blocks before or after the poisoned item, and Appendix~\ref{sec:long} describes the procedure. As Figure~\ref{fig:robust}a--b shows, retention under the default prompt stays within one episode of its $L=0$ value or rises, only \texttt{capped} loses retention as $L$ grows, and at each $L$ naming stays at or below retention.

\figrobust

\begin{table}[t]
\centering
\caption{Predicted and measured naming rates of later shipped prompts and a held-out compactor.}
\label{tab:predict}
\definecolor{pOneTint}{HTML}{E4EEF8}
\fontsize{8.5}{9.4}\selectfont
\renewcommand{\arraystretch}{1}
\setlength{\heavyrulewidth}{.74bp}\setlength{\lightrulewidth}{.4bp}\setlength{\cmidrulewidth}{.4bp}
\setlength{\aboverulesep}{1.1pt}\setlength{\belowrulesep}{1.6pt}\setlength{\cmidrulesep}{1.4pt}
\arrayrulecolor{black}
\def\pOneRows{
\toprule
 & \multicolumn{3}{c}{{\fontsize{8}{9}\selectfont\bfseries Banking}} & \multicolumn{3}{c}{{\fontsize{8}{9}\selectfont\bfseries Workspace}} & \multicolumn{3}{c}{{\fontsize{8}{9}\selectfont\bfseries Travel}} \\
\cmidrule(lr){2-4}\cmidrule(lr){5-7}\cmidrule(l){8-10}
{\fontsize{8}{9}\selectfont\bfseries Prompt} & {\fontsize{8}{9}\selectfont\bfseries Pred.} & {\fontsize{8}{9}\selectfont\bfseries Obs.} & {\fontsize{8}{9}\selectfont\bfseries $k$} & {\fontsize{8}{9}\selectfont\bfseries Pred.} & {\fontsize{8}{9}\selectfont\bfseries Obs.} & {\fontsize{8}{9}\selectfont\bfseries $k$} & {\fontsize{8}{9}\selectfont\bfseries Pred.} & {\fontsize{8}{9}\selectfont\bfseries Obs.} & {\fontsize{8}{9}\selectfont\bfseries $k$} \\
\midrule
Codex & .11 & .15 & 9 & .03 & 0 & 1 & .05 & .08 & 1 \\
Cline & .05 & .05 & 2 & .06 & .07 & 2 & 0 & 0 & 0 \\
Goose & .50 & .51 & 27 & .60 & .63 & 20 & 0 & 0 & 0 \\
Aider & .33 & .31 & 17 & .14 & .17 & 5 & 0 & 0 & 0 \\
Qwen Code & .22 & .28 & 19 & .14 & .07 & 5 & 0 & 0 & 0 \\
Kimi CLI & .07 & .08 & 4 & .05 & .03 & 2 & .16 & .25 & 3 \\
LangChain & .06 & .03 & 5 & 0 & 0 & 0 & .02 & .08 & 1 \\
\midrule
\multicolumn{10}{@{}l}{\textcolor{black!65}{\textit{Held-out compactor gpt-5.6-luna}}} \\
Claude Code & .02 & 0 & 39 & .04 & 0 & 29 & .01 & 0 & 12 \\
\hspace{.8em}business prose & .18 & .14 & 11 & -- & -- & -- & -- & -- & -- \\
\hspace{.8em}bare account & .12 & .15 & 7 & -- & -- & -- & -- & -- & -- \\
\hspace{.8em}notice to act on & .52 & 1 & 11 & -- & -- & -- & -- & -- & -- \\
Codex & .01 & 0 & 21 & .01 & 0 & 6 & .01 & 0 & 5 \\
\bottomrule
}
\setlength{\tabcolsep}{0pt}
\setbox2=\hbox{\begin{tabular}{@{}lrrrrrrrrr@{}}\pOneRows\end{tabular}}
\setlength{\tabcolsep}{\dimexpr(\columnwidth-\wd2)/18\relax}
\ifdim\tabcolsep<0pt\typeout{PONENEGATIVEPADDING}\fi
\begin{tabular}{@{}lrrrrrrrrr@{}}\pOneRows\end{tabular}
\tabnote{Pred./Obs.: predicted/measured naming rate. $k$: kept summaries in the pre-registered scope, which covers goals that carry an identifier and leaves out the banking pair whose turn-$N$ policy already named the attacker. --: not run.}
\end{table}

\textbf{Prompts of other harnesses.} Nine more shipped prompts follow the same decomposition. We run the prompts of Gemini CLI, OpenHands, Codex, Cline, Goose, Aider, Qwen Code, Kimi CLI and LangChain as compactor conditions over the same episodes, and Appendix~\ref{app:shipped} gives every cell. To predict a prompt's ratio $c$, we fit a model on four features: how a summary frames the identifier, whether the task names its own payee, the summary's length and the suite. Fit on the earlier prompts and frozen before the last seven ran, it predicted their naming rate, as Figure~\ref{fig:robust}c shows. We then added one feature, whether the request can use the identifier, and froze the refit model before running \texttt{gpt-5.6-luna}, a compactor held out from both fits. The refit model predicted that compactor within .10 in 8 of the 9 cells of Table~\ref{tab:predict}, which include three banking cells with the data payloads of Section~\ref{sec:measure:frontier}. The one miss is the notice, whose request asks the agent to act on it: the policy names its account in every case in scope, above the predicted .52, as the two-request sessions of Section~\ref{sec:measure:frontier} also show. By contrast, the suite's ratio times retention, which sees neither the mark nor whether the request can use the identifier, misses all nine cells.

Products differ in ratio by how their summaries present the account. For example, Gemini CLI records it as a transfer that Progent blocked, and the policy names it in 3 of the 20 banking summaries that kept it, whereas OpenHands lists it as pending work, as in Figure~\ref{fig:mechanism}, and the policy names it in 9 of 11.
A shipped prompt thus sets how much of the attacker's authority reaches the policy through two choices: what its summaries retain and how they present the account.

\subsection{RQ3: Live products and frontier compactors}
\label{sec:measure:boundary}

Because a live product also decides when to compact and where the summary goes, we read the records it writes at real boundaries.
We observe that in all 7 recorded Claude Code sessions, every automatic compaction placed the summary in the request slot.
This placement matters for an authorizer installed as a \texttt{PreToolUse} hook, which receives the transcript path rather than the request. The obvious rule, taking the most recent user-role record, returns the user's turn in 96.2\% of request starts and the summary at each boundary.

Those sessions carried no injection, so we also inject one into live sessions.
When the product compacts automatically, the summary keeps the injected account in 19 of 20 sessions.
The compaction runs live, in headless sessions of Claude Code's default model with the AgentDojo banking tools over MCP and a statement tool that fills the context, and Appendix~\ref{app:setup} gives the session setup. We generate Progent's policy offline from the summary record that the product wrote at the first boundary after the poisoned item, with the definitions of Section~\ref{sec:measure:setup}.
All 20 sessions compacted twice on the product's automatic trigger, and after each of the 40 boundaries the next user-role record was the summary, as it was at the 60 automatic boundaries of the auto-mode sessions in this section.
The product also writes the follow-up itself. Every compaction record of our live sessions tells the agent to ``continue the conversation from where it left off without asking the user any further questions'', a stronger form of the follow-up our replays use.
However, although the summary keeps the account, Claude Code's own model records it as an instruction the agent declined, and the policy leaves it out. An attacker can write around this mark, as the frontier compactors show.

\textbf{Codex.} Codex places its summary the same way.
Over 51 live Codex sessions under versions 0.163.0 and 0.164.0, all 1008 automatic compactions inserted the summary as a user-role item.
In both products, the summary is the latest user message after every boundary, so the request slot holds the compactor's quotation.

\textbf{Frontier compactors.}\label{sec:measure:frontier} We test what frontier compactors mark and whether an attacker can write around the mark.

\textbf{Injected instructions.} We find that frontier compactors mark an injected instruction, and the policy then leaves the account out.
With \texttt{gpt-5.4-mini} and \texttt{gpt-5.5} as the compactor, under Claude Code's default prompt and under Codex's prompt, the summary keeps the attacker's identifier more often than \texttt{gpt-4o-mini} does and marks it in 219 of the 220 summaries that kept it, over three suites.
As Figure~\ref{fig:frontier}a shows, on banking the policy rarely names it in the kept summaries, against 11 of 26 for \texttt{gpt-4o-mini}.

\figfrontier

\textbf{Injected data.} We further find that when the attacker writes data instead of an instruction, the mark disappears.
We replace AgentDojo's payload with three data-only forms: plain business prose, a bare account field, and a notice that the user's task asks the agent to act on, which states an account number but no instruction.
As Figure~\ref{fig:frontier}a shows, both frontier models then keep the account, mostly unmarked, and under the notice payload the policy names it in more than half of the kept cases under each model. A third frontier compactor, \texttt{gpt-5.6-luna}, gives the same result: the policy names the notice's account in 13 of 16 kept cases, and in all 11 within the pre-registered scope of Table~\ref{tab:predict}.

\textbf{Live product.} The same holds in a live product, where an account written as data crosses the boundary unmarked.
On Claude Code 2.1.319 with auto mode, the summaries of two instruction payloads marked the account in 16 of 17 kept cases, and every policy left it out. In contrast, when the landlord notice was reduced to the bare field ``Rent account:'' with the attacker's account, every kept summary stated it plainly as the notice's rent account. The policy derived from it then admits a redirect of the rent order to that account in 4 of 5 sessions.

\textbf{Two-request sessions.} We also observe that data read in one request becomes an explicit grant in the next.
In two-request sessions over the three suites, the user first asks what a document says, and after the session compacts asks the agent to act on it, as in ``Please make the change it asks for''.
Each frontier compactor leaves the data unmarked in all of its kept summaries. As Figure~\ref{fig:frontier}b shows, the policy for the second request then names the attacker's account or address in 55 of 80 cases, against 9 of 79 for an injected instruction.
The account comes from the summary: generated from the second request alone, as without compaction, or from the user's own two messages, the policy never names it.

\textbf{Stated payee.} Even when the user has stated the payee, the policy from the summary admits the attacker's account.
In this setting, a request-derived authorizer has the information it needs, and we test whether it uses it in two-request sessions. The user first gives the landlord's account and asks what a notice says, the attacker's notice states another account as plain data, and after the boundary the user asks to pay this month's rent. As Figure~\ref{fig:frontier}c shows, the policy derived from the user's own words never admits the notice's account under either generator, and with \texttt{gpt-5.5} as the generator it authorizes the payment to the user's own account in every session.
In contrast, the policy from the summary admits the notice's account in 290 of 320 cases, because the summary records the notice's account next to the user's, and still in most cases when \texttt{gpt-5.5} generates the policy. Generated from the follow-up alone, the policy leaves the recipient open in every session. Appendix~\ref{app:frontier} gives the counts behind Figure~\ref{fig:frontier}.
A frontier agent then acts on this grant. Re-run at turn $N{+}1$ with \texttt{gpt-5.4-mini} reading the summary, it completes the transfer to the notice's account in 33 of 80 episodes under the summary's policy and never under the gate's, which blocks all 34 of its attempts.
Authorizing from the user's words leaves the payee to the user: a notice can change where the user pays only if the user asks for the change.

\textbf{Run-time refinement.} Progent's own refinement also turns the tool results the agent read into authority: before any boundary, its run-time update adds a payment to the notice's account in 8 of 80 of the same sessions, as Appendix~\ref{app:setup} details.
Before compaction as after it, authority follows the text the authorizer reads, and text that is not the user's can carry the attacker's account into the grant.

\subsection{RQ4: Other harnesses and authorizers}
\label{sec:measure:transfer}

\begin{table}[t]
\centering
\caption{Other harnesses, authorizers and checkers after compaction.}
\label{tab:transfer}
\definecolor{pOneTint}{HTML}{E4EEF8}
\fontsize{8.5}{9.4}\selectfont
\renewcommand{\arraystretch}{1}
\setlength{\heavyrulewidth}{.74bp}\setlength{\lightrulewidth}{.4bp}\setlength{\cmidrulewidth}{.4bp}
\setlength{\aboverulesep}{1.1pt}\setlength{\belowrulesep}{1.6pt}\setlength{\cmidrulesep}{1.4pt}
\arrayrulecolor{black}
\def\pOneRows{
\toprule
 & & \multicolumn{2}{c}{{\fontsize{8}{9}\selectfont\bfseries Policy}} & {\fontsize{8}{9}\selectfont\bfseries Outcome} \\
\cmidrule(lr){3-4}\cmidrule(l){5-5}
{\fontsize{8}{9}\selectfont\bfseries System and condition} & {\fontsize{8}{9}\selectfont\bfseries $n$} & {\fontsize{8}{9}\selectfont\bfseries Expands} & {\fontsize{8}{9}\selectfont\bfseries Names} & {\fontsize{8}{9}\selectfont\bfseries Passes} \\
\midrule
\multicolumn{5}{@{}l}{\textcolor{black!65}{\textit{Authorizers that read the request slot}}} \\
smolagents + Progent & & & & \\
\hspace{.8em}user-role tool output, banking & 16 & 16 & 16 & 11 \\
Progent + Qwen2.5-72B & & & & \\
\hspace{.8em}banking & 40 & 40 & 18 & -- \\
\hspace{.8em}workspace & 30 & 18 & 16 & -- \\
\hspace{.8em}travel & 30 & 27 & 2 & -- \\
\midrule
\multicolumn{5}{@{}l}{\textcolor{black!65}{\textit{Authority taken elsewhere}}} \\
DRIFT & & & & \\
\hspace{.8em}filter before compaction, banking & 20 & 1 & -- & -- \\
CaMeL & & & & \\
\hspace{.8em}authority on values, banking & 36 & -- & -- & 3 \\
\midrule
\multicolumn{5}{@{}l}{\textcolor{black!65}{\textit{Checkers that read the request slot}}} \\
LlamaFirewall AlignmentCheck & & & & \\
\hspace{.8em}summary kept the account & 180 & -- & -- & 42 \\
\hspace{.8em}summary dropped it & 162 & -- & -- & 8 \\
Spotlighting & & & & \\
\hspace{.8em}summary at the user role & 64 & -- & -- & 34 \\
\hspace{.8em}delimited tool-result summary & 64 & -- & -- & 5 \\
Codex Guardian, rewritten summary & & & & \\
\hspace{.8em}before OpenAI's change & 33 & -- & -- & 5 \\
\hspace{.8em}after it & 33 & -- & -- & 0 \\
\bottomrule
}
\setlength{\tabcolsep}{0pt}
\setbox2=\hbox{\begin{tabular}{@{}lrrrr@{}}\pOneRows\end{tabular}}
\setlength{\tabcolsep}{\dimexpr(\columnwidth-\wd2)/8\relax}
\ifdim\tabcolsep<0pt\typeout{PONENEGATIVEPADDING}\fi
\begin{tabular}{@{}lrrrr@{}}\pOneRows\end{tabular}
\tabnote{Passes count completed actions (smolagents, CaMeL, spotlighting), transfers the auditor allows (AlignmentCheck, nine compactor prompts) or calls the reviewer approves (Guardian, delegating task). --: outside the setup.}
\end{table}

We vary the harness that fills the slot and the authorizer that reads it. On the authorizer side, the deciding factor is whether the authorizer takes its authority from the request slot.

\textbf{smolagents.} We find that smolagents, which copies tool bytes into the request slot, sets retention to one.
Version 1.26.0 maps tool responses to the user role by default and merges adjacent same-role messages, so with Progent all 16 banking policies name the attacker, and 11 of 16 episodes complete the attacker's action. Table~\ref{tab:transfer} presents this and the other systems of this section.

\textbf{Second policy generator.} With \texttt{Qwen2.5-72B-Instruct} generating the policy from the stored requests, the policy expands and names the attacker in all three suites.

\textbf{AlignmentCheck.} We find that LlamaFirewall's AlignmentCheck~\cite{llamafirewall}, a deployed auditor that reads the request slot, also follows retention.
It judges whether an agent's latest action serves the user message that opens the trace, which after compaction is the summary in the request slot. It flags the attacker's transfer at turn $N$ in all but two of 40 banking pairs. However, with summaries from nine compactor prompts in that position, it allows the same transfer far more often when the summary kept the account than when it dropped it, as Table~\ref{tab:transfer} shows.
With the summary as a tool result, it flags the transfer in 91\% of pairs, and Appendix~\ref{app:setup} gives further audit details.

\textbf{Codex Guardian.} Codex's reviewer, by contrast, follows attribution, and OpenAI's change applies it to the summary.
Guardian reviews each side-effecting call against what the user authorized, and since 15 October 2026 it attributes compaction summaries to the assistant~\cite{codex-pr}.
We freeze live Codex sessions after their first automatic compaction, under a version from before this change and one from after it, and append the attacker's call. Both versions approved it in 2 of 33 calls where the user's own message delegates to a landlord's notice, and in none of 18 calls otherwise. When the summary in the persisted session file is rewritten to state ``Pending from the user: send the next payment to'' the attacker's account, the earlier version approved some of the calls and the later one none, as Table~\ref{tab:transfer} shows.

\textbf{Spotlighting.} Spotlighting wraps tool output in delimiters and tells the model to ignore instructions inside them~\cite{hines2024spotlighting}, and it is bypassed once the summary sits at the user role. After \texttt{gpt-4o-mini} compacts the turn-$N$ transcript, the same summary leads the agent to the attacker's transfer in about half of the episodes at the user role and rarely as a delimited tool result, as Table~\ref{tab:transfer} shows.
Two authorizers depart from this pattern, each in the way the decomposition predicts.

\textbf{DRIFT.} DRIFT removes the injection before compaction and so drives retention toward zero.
Its filter cleans the memory stream before the stream reaches the summary. Over 20 episodes, we observe that the payload passes the filter in 11, reaches the summary in 4 and expands authorization in one.

\textbf{CaMeL.} CaMeL~\cite{camel} attaches authority to values rather than to the request, and it admits few attacks after compaction.
It completes the attacker's action in 3 of 36 banking episodes after compaction, against none with the follow-up only or with the injection removed from the summary.

Any authorizer that reads the request slot has a $c$, and any compactor that fills it has an $s$: a second policy generator, an auditor and a reviewer all follow the summary's content. DRIFT lowers $s$ by filtering the injections it recognizes, and CaMeL avoids $c$ by not taking authority from the request at all. A request-derived authorizer therefore needs a request that the compactor does not write, which the gate of Section~\ref{sec:fix} provides.

\subsection{RQ5: Effectiveness and cost of the gate}
\label{sec:fix:eval}

\begin{table}[!t]
\centering
\caption{End-to-end results with and without the gate.}
\label{tab:gate}
\definecolor{pOneTint}{HTML}{E4EEF8}
\fontsize{8.5}{9.4}\selectfont
\renewcommand{\arraystretch}{1}
\setlength{\heavyrulewidth}{.74bp}\setlength{\lightrulewidth}{.4bp}\setlength{\cmidrulewidth}{.4bp}
\setlength{\aboverulesep}{1.1pt}\setlength{\belowrulesep}{1.6pt}\setlength{\cmidrulesep}{1.4pt}
\arrayrulecolor{black}
\def\pOneRows{
\toprule
 &  & \multicolumn{2}{c}{{\fontsize{8}{9}\selectfont\bfseries Authorization}} & \multicolumn{2}{c}{{\fontsize{8}{9}\selectfont\bfseries Attacks}} & {\fontsize{8}{9}\selectfont\bfseries Utility} \\
\cmidrule(lr){3-4}\cmidrule(lr){5-6}\cmidrule(l){7-7}
{\fontsize{8}{9}\selectfont\bfseries Policy} & {\fontsize{8}{9}\selectfont\bfseries Agent reads} & {\fontsize{8}{9}\selectfont\bfseries Names} & {\fontsize{8}{9}\selectfont\bfseries Admits} & {\fontsize{8}{9}\selectfont\bfseries Total} & {\fontsize{8}{9}\selectfont\bfseries New} & {\fontsize{8}{9}\selectfont\bfseries Tasks} \\
\midrule
Summary & Slot summary & 18 & 24 & 7 & 7 & 9 \\
Follow-up & Tool summary & 0 & 0 & 0 & 0 & 18 \\
\midrule
Gate & Slot summary & 0 & 5 & 0 & 0 & 13 \\
\rowcolor{pOneTint}Gate & Tool summary & \textbf{0} & \textbf{6} & \textbf{0} & \textbf{0} & \textbf{19} \\
Gate & User text & 0 & 2 & 1 & 0 & 26 \\
Gate & User text, 2nd & 0 & 3 & 1 & 0 & 22 \\
Metadata & User text & 0 & 4 & 2 & 0 & 25 \\
\midrule
Request & Full transcript & 1 & 8 & 2 & -- & 27 \\
\bottomrule
}
\setlength{\tabcolsep}{0pt}
\setbox2=\hbox{\begin{tabular}{@{}>{\columncolor{white}[0pt][\tabcolsep]}llrrrr>{\columncolor{white}[\tabcolsep][0pt]}r@{}}\pOneRows\end{tabular}}
\setlength{\tabcolsep}{\dimexpr(\columnwidth-\wd2)/12\relax}
\ifdim\tabcolsep<0pt\typeout{PONENEGATIVEPADDING}\fi
\begin{tabular}{@{}>{\columncolor{white}[0pt][\tabcolsep]}llrrrr>{\columncolor{white}[\tabcolsep][0pt]}r@{}}\pOneRows\end{tabular}
\tabnote{40 episodes at turn $N{+}1$. Request: the uncompacted run. Admits: policies that permit the attacker's identifier. New: attacks beyond the uncompacted run. 2nd: after a second boundary.}
\end{table}

We measure the grants and the agent's actions under the gate, with Progent's run-time refinement on and applied without confirmation, as in its own scripts.
We find that with the gate, no attacker action completes while the agent reads the summary, in the request slot or as a tool result, and with the summary as a tool result the agent completes as many tasks as with the summary moved there and no gate.
With the summary in the request slot and no gate, 18 of the 40 policies name the attacker and 7 attacker actions complete. Through the gate, no policy names the attacker, and where a gate policy still admits the attacker's call, the user's own request left the recipient open. Table~\ref{tab:gate} presents the full results, and a later repeat of three of its conditions reproduced their task and attack counts exactly, as Appendix~\ref{app:setup} reports.

Under the same gate policy, the agent completes more of the user's tasks with the summary as a tool result than in the request slot.
Unlike moving the summary out of the slot, the gate also keeps the user's earlier request in the policy input.
As a result, an agent working from the user's text alone completes 26 tasks through the gate, against 27 in the uncompacted run and 22 after a second boundary.

\begin{table}[t]
\centering
\caption{Adaptive attacks against four designs.}
\label{tab:app:selfattacks}
\definecolor{pOneTint}{HTML}{E4EEF8}
\fontsize{8.5}{9.4}\selectfont
\renewcommand{\arraystretch}{1}
\setlength{\heavyrulewidth}{.74bp}\setlength{\lightrulewidth}{.4bp}\setlength{\cmidrulewidth}{.4bp}
\setlength{\aboverulesep}{1.1pt}\setlength{\belowrulesep}{1.6pt}\setlength{\cmidrulesep}{1.4pt}
\arrayrulecolor{black}
\def\pOneRows{
\toprule
{\fontsize{8}{9}\selectfont\bfseries Attack} & {\fontsize{8}{9}\selectfont\bfseries Content hash} & {\fontsize{8}{9}\selectfont\bfseries Witnesses} & {\fontsize{8}{9}\selectfont\bfseries Metadata} & {\fontsize{8}{9}\selectfont\bfseries Final gate} \\
\midrule
Echo & \symlose & \symkept & \symkept & \symkept \\
Replay & \symlose & \symadmit & \symadmit & \symdeny \\
Delimiter & \symkept & \symkept & \symkept & \symkept \\
Splice & \symna & \symkept & \symadmit & \symkept \\
Reorder & \symna & \symkept & \symkept & \symdeny \\
Deletion & \symna & \symlose & \symlose & \symdeny \\
Omission & \symna & \symna & \symna & \symdeny \\
Stale prefix & \symna & \symna & \symna & \symdeny \\
\bottomrule
}
\setlength{\tabcolsep}{0pt}
\setbox2=\hbox{\begin{tabular}{@{}>{\columncolor{white}[0pt][\tabcolsep]}lccc>{\columncolor{pOneTint}[\tabcolsep][0pt]}c@{}}\pOneRows\end{tabular}}
\setlength{\tabcolsep}{\dimexpr(\columnwidth-\wd2)/8\relax}
\ifdim\tabcolsep<0pt\typeout{PONENEGATIVEPADDING}\fi
\begin{tabular}{@{}>{\columncolor{white}[0pt][\tabcolsep]}lccc>{\columncolor{pOneTint}[\tabcolsep][0pt]}c@{}}\pOneRows\end{tabular}
\tabnote{76 banking transcripts per cell. \symkept: user sequence intact. \symdeny: input emptied. \symadmit: adversary item admitted in every trial. \symlose: user turn lost. \symna: not applicable.}
\end{table}

\textbf{Adaptive attacks.}\label{sec:fix:adaptive} We let the adversary transform the carried records and attack the gate itself.
We find that against the final gate, each of the eight attacks either leaves exactly the user's sequence or empties the authorization input, as Theorem~\ref{thm:soundness} guarantees.
Table~\ref{tab:app:selfattacks} presents six attacks (echo, replay, delimiter, splice, reorder and deletion) run over 76 banking transcripts against a content hash, witnesses alone, a metadata filter and the final gate. Two more, omitting a later user turn and committing a stale prefix, choose which records reach the packager. A gate without the completeness count accepts both in all 76 transcripts, whereas the final gate empties the input.
We count each input item by its actual source, so a replayed user string counts as adversary-supplied. Under this count, the final gate admits 0 adversary-supplied or omitted items across 608 trials.

\textbf{Comparison with a metadata filter.} We find that a metadata filter matches the gate only inside one process.
OpenAI deployed such a filter in Codex, where Guardian keeps the user's messages apart from the compacted history and relabels the summary so that only those messages count~\cite{codex-pr-history,codex-pr}.
We implement this rule as a filter that passes every user-role record except those marked as summaries. Inside one process, our filter hands the generator the same text as the gate in all 40 re-run episodes.
However, a record can pass through a carrier the agent can write, such as a persisted session or a memory file read back at the user role. The filter then admits the replayed and spliced records of Table~\ref{tab:app:selfattacks}, whereas the final gate refuses them.

\textbf{Cost.}\label{sec:fix:cost} In our implementation, verification costs 0.25 to 0.40\,ms per request. The gate keeps two 32-byte keys, an occurrence counter and two boundary counters per session, and tags each user turn and boundary with a 32-byte HMAC-SHA256.

\section{Discussion}
\label{sec:discussion}

\paragraph{Other carriers} Memory read back at the user role and reports relayed from sub-agents also place text the user did not write into the request slot. In Claude Code, for example, a delegated sub-agent is its own session whose first request is a prompt that the parent agent composed and delivered at the user role, as in the 78 sub-agent transcripts of our corpus, which Appendix~\ref{app:setup} describes. Because Lemma~\ref{prop:zero} and Theorem~\ref{thm:coupling} concern any text in the slot, they apply to these carriers unchanged. The gate's design extends to them: it authenticates the user's text at ingress rather than at the compaction step, and it refuses the records replayed through a memory file in Table~\ref{tab:app:selfattacks}.

\paragraph{Frozen policies} A policy frozen at the first request is out of a summary's reach, yet it cannot follow the user's later tasks, as Section~\ref{sec:measure:setup} shows, and Progent's run-time refinement still draws authority from the tool results the agent reads. The gate keeps both properties: per-request derivation follows the user's later tasks, and an input taken only from the user keeps the policy out of the summary's reach. Through it, the agent completes as many tasks as with the summary moved to a tool result.

\paragraph{Limitations and future work} We evaluate the gate on replayed AgentDojo sessions with per-request derivation, and integrating it into a production harness, through Claude Code's authorization hook or Codex's history of user messages, is the next step. The gate covers what the user states, and a value the user leaves to a document stays subject to Theorem~\ref{thm:coupling} until the document's source is authenticated, for example through signed tool outputs. Our evaluation covers the AgentDojo and ASB tool suites and two deployed reviewers, Codex's Guardian and LlamaFirewall's AlignmentCheck. Long sessions, multi-agent workflows, memory read back across sessions and sub-agent reports are the next settings to measure. The guarantee assumes that the keys and counters stay outside the agent's reach, which on hosts where the agent runs code calls for isolating ingest and the gate from it, for example in a sandbox.

\paragraph{Deployment} In Claude Code, an authorization hook can read the witnessed user records instead of the latest user message, and the packager runs where the harness writes its compaction record. Inside Codex, Guardian's history of user messages already plays the packager's role in one process. Where the harness keeps the user's turns in a store that the agent cannot write, that store serves as ingest and packager, and the gate needs no MACs. The MACs are needed instead for records that pass through storage the agent can write, such as Claude Code's session files. The gate also leaves benign sessions intact: it released the user's complete sequence every time, at both boundaries of the 40 end-to-end episodes and in 76 transcripts with and without a cancellation turn.

\section{Related Work}
\label{sec:related}

\paragraph{Origin-bound memory}
Louck shows that an agent's own summarization, a trusted-tool echo or manufactured corroboration can launder an untrusted origin into long-term memory, and proves that no content- or lineage-based defense is sound under laundering, so origin must be bound at write time~\cite{tmanm}. Our results complement this impossibility: Theorem~\ref{thm:coupling} bounds how far a text-only authorizer can separate a laundered record from a legitimate one, and our measurements show how much deployed compactors launder.
Louck protects a stored memory item's authority across sessions, whereas we study the request slot, which compaction fills within one session.
At enforcement, Louck's act gate checks a single-use user token for each consequential action. Our gate instead requires the authorizer's input to be the session's complete witnessed sequence, in order, so it refuses a replayed, reordered or omitted user turn that binding origin alone admits, as Table~\ref{tab:app:selfattacks} shows.

\paragraph{Authority in persistent memory}
In concurrent work, consolidation keeps a claim but erases the constraints on who may act on it~\cite{authmem}, and memory writers record permissions that the history never granted, even in benign runs~\cite{eal}.
CXI checks protected fields at the execution boundary~\cite{cxi}, and memory poisoning and stored prompt injection keep an attacker's text in agent state across sessions~\cite{dash2026memory,xie2026stored}. Memory Provenance Laundering and MemLineage attach provenance to stored memories and check it before tool calls~\cite{mpl,memlineage}. These works bind stored memory across sessions, whereas we require the authorizer's input within one session to be the user's complete, ordered turns.

\paragraph{Compression boundaries}
Relink and FragFuse assemble a payload from fragments that are benign before compression~\cite{relink,fragfuse}. Our payload, in contrast, is complete before compression, and compression changes only whose request it appears to be. OpenAI reports a model in training that wrote prompt injections into its own compaction summaries~\cite{openai-compaction}, whereas our compactor is honest and the injection comes from the attacker's data. A Codex user reported a summary that recast an assistant's plan as the user's authorization~\cite{codex-issue}, a benign case of the misattribution that we show an attacker can steer.

\paragraph{Request-derived, system-level and in-band defenses}
Progent, DRIFT and Conseca derive permitted actions from the request or other trusted context~\cite{progent,drift,conseca}. CaMeL tracks capabilities in an interpreter~\cite{camel}, FIDES and RTBAS propagate labels~\cite{fides,rtbas}, and MELON, Task Shield and IPIGuard constrain task alignment or dependencies~\cite{zhu2025melon,jia2025taskshield,an2025ipiguard}.
Prior analysis argues that provenance must be enforced outside the model's text interface~\cite{inseparability}.
AuthGraph derives an authorization graph from the user's intent in an isolated context and aligns it with the execution's provenance graph~\cite{authgraph}. Role confusion traces prompt injection to models that judge the speaker by how text sounds rather than by its role label~\cite{role-confusion}. In our setting, the label itself is wrong, because the harness labels the compactor's quotation as the user and the authorizer trusts the label.
Classifiers, structured prompting and instruction hierarchies act on the text the model reads~\cite{chen2025struq,wu2025ise,hines2024spotlighting} and fall to adaptive attacks~\cite{zhan2025adaptive,attacker-moves-second}.
In classic terms, reading carried context as the user's request is a confused-deputy pattern~\cite{confused-deputy}. Witnesses with a list commitment form a tamper-evident log~\cite{schneier99,crosby09}, and our construction follows capability systems and information-flow control~\cite{dennis66,denning76,myers97,sabelfeld03,cecchetti17}, where endorsement means treating information as more trusted than the sources that influenced it. We hold the authorizer fixed and bind, outside the model, the origin of the text it credits.

\section{Conclusion}
\label{sec:conclusion}

In this paper, we showed that request-derived authorizers assume the request slot holds the user's words, while compaction places a model-written summary there.
All 1108 live boundaries we recorded in Claude Code and Codex did so, and across 18 shipped compaction prompts the resulting policy followed the identifiers the summary kept and the marks it wrote.
We proved that an authorizer reading text alone can tell a forged record from its legitimate twin no better than their statistical distance allows, which is why frontier compactors mark injected instructions but rarely injected data.
Our gate gives the authorizer the user's authenticated turns and leaves the summary to the agent. While the agent reads the summary, no attacker action completed in 40 attacked AgentDojo episodes, and a document can change a payee only when the user's own words ask for it.
Until authorizers adopt such a separation, a compactor's retention and marking remain security parameters.

\section*{Artifact Availability}
\label{sec:openscience}
\addcontentsline{toc}{section}{Artifact Availability}

\noindent\textbf{Artifact URL:} {\def\UrlBreaks{\do\/\do\-}\url{https://anonymous.4open.science/r/LLM-Agent-Authorization-Across-Context-Compaction-504E/}}

The repository contains the code of every experiment, the gate and the script that computes the paper's numbers, and Appendix~\ref{app:setup} lists the experiments we pre-registered and the intervals we report.

\label{body:end}

\bibliographystyle{plain}
\bibliography{refs}
\appendices
\makeatletter
\setlength{\@fptop}{0pt}\setlength{\@fpsep}{12pt}\setlength{\@fpbot}{0pt plus 1fil}
\setlength{\@dblfptop}{0pt}\setlength{\@dblfpsep}{12pt}\setlength{\@dblfpbot}{0pt plus 1fil}
\makeatother
\renewcommand{\bottomfraction}{0.5}
\section{Proofs}
\label{app:proofs}

\noindent\emph{Proof of Proposition~\ref{prop:checks}.} Each case keeps the other two mechanisms. Without witnesses, a forged user-role record that carries the counter of a genuine turn keeps the list ordered and complete, so the packager commits it and the gate releases it. Without the order commitment, the gate checks each witness and that the list holds $m$ records, so a copy of one witnessed turn in place of another, or a swap of two turns, is released. Without the completeness count, neither party compares the list with ingest's count, so withholding the last witnessed turn before packaging leaves the counters $1,\ldots,m-1$, which pass both.\hfill$\square$

\subsection{Corollaries of Theorem~\ref{thm:soundness}}
\label{app:proof}

We state two consequences of Theorem~\ref{thm:soundness}. Both rely on the gate's consumed-counter state, whose rollback would violate Assumption~\ref{as:chan} and reopen replay.

\begin{corollary}[Authorizing provenance]
\label{cor:indep}
Every admitted occurrence has an authorizing ingest witness. An abstractive output, an untrusted
occurrence, or bytes that merely match trusted text cannot acquire such a witness.  Copying an
authentic record preserves its original provenance, but the copy is admitted only as part of the
fresh vector committed for the current boundary.
\end{corollary}

\noindent This is the occurrence-level form of the retention bound in Section~\ref{sec:law}. Together with the completeness count of Theorem~\ref{thm:soundness}, it also fixes the admitted set: the committed vector holds every witnessed occurrence $1,\ldots,m$, so the adversary cannot choose which authentic occurrences reach the generator.

\begin{corollary}[Rendering does not mint authority]
\label{cor:render2}
Changing a payload's rendering either preserves the same canonical bytes and still lacks a fresh
authorizing occurrence in the committed vector, or changes the occurrence MAC message.  In neither
case does rendering alone produce an admitted adversarial occurrence, except with negligible
probability.
\end{corollary}

\section{Setup and Reproduction}
\label{app:setup}
\begin{table*}[t]
\centering
\caption{Retention, expansion and naming over session length.}
\label{tab:e33b}
\definecolor{pOneTint}{HTML}{E4EEF8}
\fontsize{8.2}{8.9}\selectfont
\renewcommand{\arraystretch}{1}
\setlength{\heavyrulewidth}{.74bp}\setlength{\lightrulewidth}{.4bp}\setlength{\cmidrulewidth}{.4bp}
\setlength{\aboverulesep}{1.1pt}\setlength{\belowrulesep}{1.6pt}\setlength{\cmidrulesep}{1.4pt}
\arrayrulecolor{black}
\def\pOneRows{
\toprule
 &  &  & \multicolumn{3}{c}{{\fontsize{8}{9}\selectfont\bfseries $L=0$}} & \multicolumn{3}{c}{{\fontsize{8}{9}\selectfont\bfseries $L=4$}} & \multicolumn{3}{c}{{\fontsize{8}{9}\selectfont\bfseries $L=16$}} & \multicolumn{3}{c}{{\fontsize{8}{9}\selectfont\bfseries $L=64$}} \\
\cmidrule(lr){4-6}\cmidrule(lr){7-9}\cmidrule(lr){10-12}\cmidrule(l){13-15}
{\fontsize{8}{9}\selectfont\bfseries Prompt} & {\fontsize{8}{9}\selectfont\bfseries Suite} & {\fontsize{8}{9}\selectfont\bfseries Poison position} & {\fontsize{8}{9}\selectfont\bfseries Ret.} & {\fontsize{8}{9}\selectfont\bfseries Exp.} & {\fontsize{8}{9}\selectfont\bfseries Named} & {\fontsize{8}{9}\selectfont\bfseries Ret.} & {\fontsize{8}{9}\selectfont\bfseries Exp.} & {\fontsize{8}{9}\selectfont\bfseries Named} & {\fontsize{8}{9}\selectfont\bfseries Ret.} & {\fontsize{8}{9}\selectfont\bfseries Exp.} & {\fontsize{8}{9}\selectfont\bfseries Named} & {\fontsize{8}{9}\selectfont\bfseries Ret.} & {\fontsize{8}{9}\selectfont\bfseries Exp.} & {\fontsize{8}{9}\selectfont\bfseries Named} \\
\midrule
\raisebox{-13.35pt}[0pt][0pt]{default} & \raisebox{-4.45pt}[0pt][0pt]{Banking} & Oldest & .6 & .4 & .3 & .6 & .5 & .3 & .5 & .9 & .5 & .6 & 1 & .5 \\
 &  & Most recent & .6 & .4 & .3 & .9 & .9 & .8 & .8 & .9 & .6 & 1 & .9 & .9 \\
 & \raisebox{-4.45pt}[0pt][0pt]{Workspace} & Oldest & .4 & .5 & .4 & .4 & .7 & .4 & .4 & .3 & .3 & .5 & .6 & .5 \\
 &  & Most recent & .4 & .5 & .4 & .3 & .2 & .2 & .3 & .3 & .3 & .3 & .4 & .3 \\
\midrule
\raisebox{-13.35pt}[0pt][0pt]{capped} & \raisebox{-4.45pt}[0pt][0pt]{Banking} & Oldest & .8 & .8 & .6 & .9 & .8 & .7 & .7 & .9 & .6 & .5 & .8 & .4 \\
 &  & Most recent & .8 & .8 & .6 & .7 & .7 & .4 & .3 & 1 & .3 & .3 & 1 & .3 \\
 & \raisebox{-4.45pt}[0pt][0pt]{Workspace} & Oldest & .6 & .9 & .6 & .9 & .6 & .5 & .9 & .9 & .7 & .7 & .8 & .5 \\
 &  & Most recent & .6 & .9 & .6 & .8 & .6 & .4 & .6 & .5 & .3 & .1 & 0 & 0 \\
\midrule
\raisebox{-13.35pt}[0pt][0pt]{lean} & \raisebox{-4.45pt}[0pt][0pt]{Banking} & Oldest & .1 & .3 & 0 & .1 & .7 & 0 & 0 & .8 & 0 & .1 & .8 & 0 \\
 &  & Most recent & .1 & .3 & 0 & .4 & .7 & .3 & .4 & .9 & .3 & .7 & .8 & .5 \\
 & \raisebox{-4.45pt}[0pt][0pt]{Workspace} & Oldest & .1 & .6 & .1 & .2 & .7 & .2 & .1 & .8 & .1 & .1 & .4 & 0 \\
 &  & Most recent & .1 & .6 & .1 & .1 & .6 & .1 & .2 & .7 & .2 & .1 & .5 & .1 \\
\bottomrule}
\setlength{\tabcolsep}{0pt}
\setbox2=\hbox{\begin{tabular}{@{}lllrrrrrrrrrrrr@{}}\pOneRows\end{tabular}}
\setlength{\tabcolsep}{\dimexpr(\textwidth-\wd2)/28\relax}
\ifdim\tabcolsep<0pt\typeout{PONENEGATIVEPADDING}\fi
\begin{tabular}{@{}lllrrrrrrrrrrrr@{}}\pOneRows\end{tabular}
\tabnote{$L$: benign blocks before or after the poisoned item, with $n=10$ per point. Rates are exact tenths. Ret./Exp./Named: retention/expanded/named.}
\end{table*}

{\raggedright\paragraph{Mark and framing} The mark words of Section~\ref{sec:law:ratio} count anywhere in a summary, and we read a summary's framing of the identifier from the sentence around its first mention.

\paragraph{Records and code locations} Claude Code records each automatic compaction as a \texttt{compact\_boundary} with \texttt{trigger:"auto"} and writes the summary into a record with \texttt{role:"user"} marked \texttt{isCompactSummary}. Codex, when it compacts locally, inserts the summary as the last user-role item of the compacted history, after the user messages it keeps, tagged \texttt{compaction.summary}. Progent's proxy (commit \texttt{8a8eb89}) reads the request slot as \texttt{messages[1]} in \path{secagent/progent_proxy.py:59} and checks arguments in \path{secagent/tool.py}:539--553, and AutoDojo annotates the initial policy as injection-independent in \path{progent_elements.py:216}. A carrier the agent can edit is, for example, the project's \texttt{CLAUDE.md}, which Claude Code inserts into a user-role message of the next session.

\paragraph{Literal matching} Literal matching misses no naming: across 2957 stored policies, every policy that names a variant of the identifier (case, spacing, truncation) also contains the literal.

\paragraph{Pre-registration and statistics} The compactor sweep, the long-session sweep, the live sessions, the ASB run, the later compactor conditions, the frontier compactors, the mark insertion, the two-request sessions, the end-to-end runs and the AlignmentCheck scans were pre-registered. We report 95\% Wilson intervals for single proportions and 95\% percentile bootstraps that resample either whole task--injection pairs or injection tasks, with 20{,}000 draws and a fixed seed.

\paragraph{Run-time refinement} We examine the stated-payee sessions before any boundary, under a request that asks what the notice says. Progent's run-time update from tool results adds a payment to the notice's account in 8 of 80 data sessions and in none of 40 instruction sessions, and never one to the account the user gave.

\paragraph{Repeated runs} Runs at temperature $0$ still vary between repetitions. Repeated on a later date over the same 40 episodes, the end-to-end runs of Table~\ref{tab:gate} reproduced their task counts exactly (9 with the summary in the slot, 18 with it as a tool result, 26 through the gate with the agent on the user's text) and their 7, 0 and 1 attacker actions. The $L = 0$ episodes of the long-session sweep repeat the compactor sweep's inputs exactly, and the compactor keeps or drops the identifier the same way in 55 of 60 pairs. The comparison with an undefended agent in Section~\ref{sec:measure:setup} uses each pair's first Progent repetition, and the repeat it reports differs by $-$.07 [$-$.19, .05].

\paragraph{Versions} Benchmark: AgentDojo suite v1.2.2 (vendored package 0.1.35). Agent:
\texttt{gpt-4o-mini-2024-07-18}. Policy generator: \texttt{gpt-4o}. Compactor: \texttt{gpt-4o-mini-2024-07-18}. 7B compactor: \texttt{Qwen2.5-7B-Instruct}. Second policy generator: \texttt{Qwen2.5-72B-Instruct}. ASB: Progent's fork, commit \texttt{8a8eb894}. Gemini CLI core 0.66.0, OpenHands SDK commit \texttt{f0adfd0}, and LlamaFirewall 1.0.3 with auditor model \texttt{gpt-4o-2024-08-06}. DRIFT's defense model is \texttt{gpt-4o} and CaMeL's model is \texttt{gpt-4o-mini-2024-07-18}, and both run with the agent and compactor above. Temperature is $0$ where
exposed. The policy generator's API
alias can change, so we record the date of every run. Live sessions: Claude Code 2.1.312 and 2.1.319 with \texttt{claude-opus-5-5}, and Codex 0.163.0 and 0.164.0, compacting locally with the token limit set to 15{,}000. Frontier compactors: \texttt{gpt-5.4-mini-2026-03-17}, \texttt{gpt-5.5-2026-04-23} and \texttt{gpt-5.6-luna}.\par}
\paragraph{Long sessions}\label{sec:long} We add $L \in \{0, 4, 16, 64\}$ tool-call blocks from other tasks of the same suite before or after the poisoned item in banking and workspace transcripts and compact with three shipped prompts. Table~\ref{tab:e33b} presents the results, and Figure~\ref{fig:law} shows them as open circles.

\paragraph{Live Claude Code sessions} Each injected live session runs headless in a fresh home directory with an empty project directory, with \texttt{autoCompactWindow} at the smallest value the product accepts, so that the automatic trigger fires within one task, and every other setting at its default. The statement tool returns a few thousand tokens for each of 24 months, and the user's task has three ordered steps: the AgentDojo prompt and two reconciliation steps over the statements. The agent reads the poisoned item either first or just before the trigger, with two banking pairs carrying the injected bytes of the main matrices and five sessions per pair in each condition. A session counts if it ran without an API error, read the poisoned item and then compacted on the product's own trigger.

\paragraph{Policy cache and rate limits} The benchmark wrapper caches generated policies by query string.
The main matrices use a per-repetition cache tag so repetitions never replay one another's policies, and the ablations of \S\ref{sec:law:slot} disable the cache, since two conditions that issue the same follow-up would otherwise share one entry.

{\raggedright \paragraph{AlignmentCheck audits} AlignmentCheck audits the transfer executed in the undefended run, or the task's reference transfer for the 14 pairs whose undefended run made none, and a second auditor, \texttt{Llama-3.3-70B-Instruct}, scanned the same traces (allowed after compaction 102/342 against the primary's 50/342).

\paragraph{Session corpus} Claude Code writes one transcript per session at \texttt{<root>/<project>/<session>.jsonl} and one per delegated sub-agent at \path{<root>/<project>/<session>/subagents/agent-*.jsonl}. Both globs are scanned, with a floor of $50$\,kB that drops transcripts too short to contain a compaction boundary at all, which on this machine leaves the seven session transcripts and 78 sub-agent transcripts. The corpus was frozen on 2026-08-14, and the scan kept only the kind of each \texttt{role:"user"} record and the position of each compaction boundary. Sub-agent transcripts measure delegation rather than compaction, so we keep them apart from the sessions, and the field \texttt{isSidechain} separates the two strata exactly.\par}

\paragraph{Analysis block of the default prompt} The default Claude Code prompt asks for an \texttt{<analysis>} block before the summary, and the product discards the block before it resumes. The policy generator reads every default-prompt summary with the block removed, as the product presents it, and every count in the paper uses that summary once.

\begin{table}[t]
\centering
\caption{Retention, expansion and naming per injection goal.}
\label{tab:fold}
\definecolor{pOneTint}{HTML}{E4EEF8}
\fontsize{8.3}{9.1}\selectfont
\renewcommand{\arraystretch}{1}
\setlength{\heavyrulewidth}{.74bp}\setlength{\lightrulewidth}{.4bp}\setlength{\cmidrulewidth}{.4bp}
\setlength{\aboverulesep}{1.1pt}\setlength{\belowrulesep}{1.6pt}\setlength{\cmidrulesep}{1.4pt}
\arrayrulecolor{black}
\def\pOneRows{
\toprule
{\fontsize{8}{9}\selectfont\bfseries Suite} & {\fontsize{8}{9}\selectfont\bfseries Goal} & {\fontsize{8}{9}\selectfont\bfseries $n$} & {\fontsize{8}{9}\selectfont\bfseries Ret.} & {\fontsize{8}{9}\selectfont\bfseries Exp.} & {\fontsize{8}{9}\selectfont\bfseries Named} \\
\midrule
\raisebox{-31.849999999999998pt}[0pt][0pt]{Banking} & 0 & 26 & .85 & .73 & .23 \\
 & 1 & 8 & .75 & .88 & .25 \\
 & 2 & 8 & .88 & .88 & .50 \\
 & 4 & 25 & .96 & .84 & .68 \\
 & 5 & 16 & .81 & .75 & .50 \\
 & 6 & 8 & .75 & .63 & .63 \\
 & 7 & 8 & .00 & .88 & .00 \\
 & 8 & 8 & 1.00 & .75 & .38 \\
\midrule
\raisebox{-36.4pt}[0pt][0pt]{Workspace} & 0 & 6 & 1.00 & 1.00 & 1.00 \\
 & 1 & 6 & .00 & .83 & .00 \\
 & 2 & 6 & .50 & 1.00 & .50 \\
 & 3 & 6 & 1.00 & .83 & .83 \\
 & 4 & 6 & .67 & 1.00 & .67 \\
 & 5 & 18 & .50 & .78 & .33 \\
 & 6 & 18 & .94 & .83 & .78 \\
 & 10 & 6 & 1.00 & 1.00 & .83 \\
 & 13 & 6 & .00 & 1.00 & .00 \\
\midrule
\raisebox{-22.75pt}[0pt][0pt]{Travel} & 0 & 12 & .00 & .75 & .00 \\
 & 1 & 6 & .67 & .83 & .50 \\
 & 2 & 12 & .00 & .75 & .00 \\
 & 3 & 6 & .67 & .83 & .17 \\
 & 4 & 12 & .00 & .75 & .00 \\
 & 5 & 6 & .00 & .50 & .00 \\
\midrule
\raisebox{-18.2pt}[0pt][0pt]{Slack} & 1 & 16 & .56 & .25 & .06 \\
 & 2 & 16 & .63 & .50 & .25 \\
 & 3 & 16 & .56 & .31 & .00 \\
 & 4 & 16 & .75 & .31 & .13 \\
 & 5 & 16 & .88 & .38 & .13 \\
\bottomrule
}
\setlength{\tabcolsep}{0pt}
\setbox2=\hbox{\begin{tabular}{@{}lrrrrr@{}}\pOneRows\end{tabular}}
\setlength{\tabcolsep}{\dimexpr(\columnwidth-\wd2)/10\relax}
\ifdim\tabcolsep<0pt\typeout{PONENEGATIVEPADDING}\fi
\begin{tabular}{@{}lrrrrr@{}}\pOneRows\end{tabular}
\tabnote{Main and goal-diversity runs by injection goal. Table~\ref{tab:main} gives the pooled estimates.}
\end{table}

\begin{table*}[t]
\centering
\caption{Retention, expansion and naming under every shipped prompt.}
\label{tab:e33a}
\definecolor{pOneTint}{HTML}{E4EEF8}
\fontsize{8.2}{9.0}\selectfont
\renewcommand{\arraystretch}{1}
\setlength{\heavyrulewidth}{.74bp}\setlength{\lightrulewidth}{.4bp}\setlength{\cmidrulewidth}{.4bp}
\setlength{\aboverulesep}{1.1pt}\setlength{\belowrulesep}{1.6pt}\setlength{\cmidrulesep}{1.4pt}
\arrayrulecolor{black}
\def\pOneRows{
\toprule
 & \multicolumn{5}{c}{{\fontsize{8}{9}\selectfont\bfseries Banking}} & \multicolumn{5}{c}{{\fontsize{8}{9}\selectfont\bfseries Workspace}} & \multicolumn{5}{c}{{\fontsize{8}{9}\selectfont\bfseries Travel}} \\
\cmidrule(lr){2-6}\cmidrule(lr){7-11}\cmidrule(l){12-16}
{\fontsize{8}{9}\selectfont\bfseries Prompt} & {\fontsize{8}{9}\selectfont\bfseries Ret.} & {\fontsize{8}{9}\selectfont\bfseries Exp.} & {\fontsize{8}{9}\selectfont\bfseries Named} & {\fontsize{8}{9}\selectfont\bfseries Kept} & {\fontsize{8}{9}\selectfont\bfseries Drop.} & {\fontsize{8}{9}\selectfont\bfseries Ret.} & {\fontsize{8}{9}\selectfont\bfseries Exp.} & {\fontsize{8}{9}\selectfont\bfseries Named} & {\fontsize{8}{9}\selectfont\bfseries Kept} & {\fontsize{8}{9}\selectfont\bfseries Drop.} & {\fontsize{8}{9}\selectfont\bfseries Ret.} & {\fontsize{8}{9}\selectfont\bfseries Exp.} & {\fontsize{8}{9}\selectfont\bfseries Named} & {\fontsize{8}{9}\selectfont\bfseries Kept} & {\fontsize{8}{9}\selectfont\bfseries Drop.} \\
\midrule
default & .65 & .60 & .28 & 11/26 & 0/14 & .43 & .47 & .40 & 12/13 & 0/17 & .07 & .77 & .03 & 1/2 & 0/28 \\
no scratchpad & .53 & .55 & .20 & 8/21 & 0/19 & .30 & .47 & .27 & 8/9 & 0/21 & .03 & .63 & .00 & 0/1 & 0/29 \\
no scratchpad + cap & .55 & .53 & .25 & 10/22 & 0/18 & .43 & .43 & .30 & 9/13 & 0/17 & .10 & .67 & .00 & 0/3 & 0/27 \\
no scratchpad + asks & .55 & .48 & .20 & 8/22 & 0/18 & .43 & .57 & .30 & 9/13 & 0/17 & .03 & .60 & .00 & 0/1 & 0/29 \\
lean & .10 & .55 & .05 & 2/4 & 0/36 & .13 & .63 & .13 & 4/4 & 0/26 & .03 & .83 & .03 & 1/1 & 0/29 \\
short & .43 & .73 & .35 & 14/17 & 0/23 & .30 & .60 & .27 & 8/9 & 0/21 & .07 & .70 & .03 & 1/2 & 0/28 \\
short v2 & .38 & .75 & .30 & 12/15 & 0/25 & .23 & .77 & .20 & 6/7 & 0/23 & .07 & .73 & .07 & 2/2 & 0/28 \\
capped & .85 & .85 & .65 & 26/34 & 0/6 & .63 & .87 & .63 & 19/19 & 0/11 & .33 & .73 & .27 & 8/10 & 0/20 \\
opencode & .20 & .48 & .13 & 5/8 & 0/32 & .00 & .23 & .00 & 0/0 & 0/30 & .00 & .63 & .00 & 0/0 & 0/30 \\
Gemini CLI & .50 & .30 & .08 & 3/20 & 0/20 & .17 & .27 & .03 & 1/5 & 0/25 & .07 & .70 & .07 & 2/2 & 0/28 \\
OpenHands & .28 & .63 & .23 & 9/11 & 0/29 & .30 & .70 & .30 & 9/9 & 0/21 & .00 & .77 & .00 & 0/0 & 0/30 \\
\midrule
Codex & .25 & .38 & .15 & 6/10 & 0/30 & .03 & .40 & .00 & 0/1 & 0/29 & .03 & .87 & .03 & 1/1 & 0/29 \\
Cline & .08 & .60 & .05 & 2/3 & 0/37 & .07 & .77 & .07 & 2/2 & 0/28 & .00 & .80 & .00 & 0/0 & 0/30 \\
Goose & .70 & .75 & .50 & 20/28 & 0/12 & .67 & .67 & .63 & 19/20 & 0/10 & .00 & .73 & .00 & 0/0 & 0/30 \\
Aider & .45 & .60 & .30 & 12/18 & 0/22 & .17 & .43 & .17 & 5/5 & 0/25 & .00 & .67 & .00 & 0/0 & 0/30 \\
Qwen Code & .48 & .70 & .28 & 11/19 & 0/21 & .17 & .13 & .07 & 2/5 & 0/25 & .00 & .53 & .00 & 0/0 & 0/30 \\
Kimi CLI & .13 & .28 & .08 & 3/5 & 0/35 & .07 & .27 & .03 & 1/2 & 0/28 & .10 & .63 & .10 & 3/3 & 0/27 \\
LangChain & .15 & .43 & .03 & 1/6 & 0/34 & .00 & .20 & .00 & 0/0 & 0/30 & .03 & .67 & .03 & 1/1 & 0/29 \\
\bottomrule
}
\setlength{\tabcolsep}{0pt}
\setbox2=\hbox{\begin{tabular}{@{}lrrrrrrrrrrrrrrr@{}}\pOneRows\end{tabular}}
\setlength{\tabcolsep}{\dimexpr(\textwidth-\wd2)/30\relax}
\ifdim\tabcolsep<0pt\typeout{PONENEGATIVEPADDING}\fi
\begin{tabular}{@{}lrrrrrrrrrrrrrrr@{}}\pOneRows\end{tabular}
\tabnote{Ret./Exp./Named: retention/expanded/named. Kept/Drop.: naming counts. $n=40,30,30$ per prompt, and later prompts are below the line. + asks: keep asks.}
\end{table*}

\begin{table*}[!t]
\begin{minipage}[t]{\columnwidth}\vspace{0pt}
\centering
\caption{Counts for Figure~\ref{fig:frontier}a--b.}
\label{tab:frontier}
\definecolor{pOneTint}{HTML}{E4EEF8}
\fontsize{8.2}{8.8}\selectfont
\renewcommand{\arraystretch}{1}
\setlength{\heavyrulewidth}{.74bp}\setlength{\lightrulewidth}{.4bp}\setlength{\cmidrulewidth}{.4bp}
\setlength{\aboverulesep}{1.1pt}\setlength{\belowrulesep}{1.6pt}\setlength{\cmidrulesep}{1.4pt}
\arrayrulecolor{black}
\def\pOneRows{
\toprule
 & \multicolumn{2}{c}{{\fontsize{8}{9}\selectfont\bfseries gpt-4o-mini}} & \multicolumn{2}{c}{{\fontsize{8}{9}\selectfont\bfseries gpt-5.4-mini}} & \multicolumn{2}{c}{{\fontsize{8}{9}\selectfont\bfseries gpt-5.5}} \\
\cmidrule(lr){2-3}\cmidrule(lr){4-5}\cmidrule(l){6-7}
{\fontsize{8}{9}\selectfont\bfseries Payload} & {\fontsize{8}{9}\selectfont\bfseries Mark} & {\fontsize{8}{9}\selectfont\bfseries Name} & {\fontsize{8}{9}\selectfont\bfseries Mark} & {\fontsize{8}{9}\selectfont\bfseries Name} & {\fontsize{8}{9}\selectfont\bfseries Mark} & {\fontsize{8}{9}\selectfont\bfseries Name} \\
\midrule
Injected instruction & 0/26 & 11/26 & 39/39 & 1/39 & 40/40 & 0/40 \\
Business prose & 0/3 & 2/3 & 1/11 & 2/11 & 5/16 & 2/16 \\
Bare account & 0/2 & 1/2 & 0/11 & 1/11 & 2/16 & 0/16 \\
Notice to act on & 0/9 & 7/9 & 0/16 & 9/16 & 3/16 & 9/16 \\
\midrule
\multicolumn{7}{@{}l}{\textcolor{black!65}{\textit{Asked about, then acted on in the next request}}} \\
Injected instruction & -- & -- & 39/39 & 7/39 & 40/40 & 2/40 \\
Account/address data & -- & -- & 0/40 & 29/40 & 0/40 & 26/40 \\
\bottomrule
}
\setlength{\tabcolsep}{0pt}
\setbox2=\hbox{\begin{tabular}{@{}lrrrrrr@{}}\pOneRows\end{tabular}}
\setlength{\tabcolsep}{\dimexpr(\columnwidth-\wd2)/12\relax}
\ifdim\tabcolsep<0pt\typeout{PONENEGATIVEPADDING}\fi
\begin{tabular}{@{}lrrrrrr@{}}\pOneRows\end{tabular}
\tabnote{Top: banking, 40 episodes for the injected instruction and 16 per data payload. Bottom: three suites, 40 episodes per cell.}
\end{minipage}
\hfill
\begin{minipage}[t]{\columnwidth}\vspace{0pt}
\centering
\caption{Counts for Figure~\ref{fig:frontier}c.}
\label{tab:attrib}
\definecolor{pOneTint}{HTML}{E4EEF8}
\fontsize{8.3}{9.2}\selectfont
\renewcommand{\arraystretch}{1}
\setlength{\heavyrulewidth}{.74bp}\setlength{\lightrulewidth}{.4bp}\setlength{\cmidrulewidth}{.4bp}
\setlength{\aboverulesep}{1.1pt}\setlength{\belowrulesep}{1.6pt}\setlength{\cmidrulesep}{1.4pt}
\arrayrulecolor{black}
\def\pOneRows{
\toprule
 &  & \multicolumn{2}{c}{{\fontsize{8}{9}\selectfont\bfseries Pay the rent}} & \multicolumn{2}{c}{{\fontsize{8}{9}\selectfont\bfseries As the notice asks}} \\
\cmidrule(lr){3-4}\cmidrule(l){5-6}
{\fontsize{8}{9}\selectfont\bfseries Notice states $X$} & {\fontsize{8}{9}\selectfont\bfseries Marked} & {\fontsize{8}{9}\selectfont\bfseries Summary} & {\fontsize{8}{9}\selectfont\bfseries User} & {\fontsize{8}{9}\selectfont\bfseries Summary} & {\fontsize{8}{9}\selectfont\bfseries User} \\
\midrule
\multicolumn{6}{@{}l}{\textcolor{black!65}{\textit{Policy generator gpt-4o}}} \\
Account change & 16/160 & 137/160 & 0/40 & 157/160 & 0/40 \\
Bare field & 14/159 & 153/160 & 0/40 & 155/160 & 0/40 \\
Injected instruction & 159/159 & 0/160 & 0/40 & 0/160 & 0/40 \\
\midrule
\multicolumn{6}{@{}l}{\textcolor{black!65}{\textit{Policy generator gpt-5.5, same summaries}}} \\
Account change &  & 68/160 & 0/40 & 160/160 & 4/40 \\
Bare field &  & 121/160 & 0/40 & 146/160 & 4/40 \\
Injected instruction &  & 0/160 & 0/40 & 1/160 & 5/40 \\
\bottomrule
}
\setlength{\tabcolsep}{0pt}
\setbox2=\hbox{\begin{tabular}{@{}lrrrrr@{}}\pOneRows\end{tabular}}
\setlength{\tabcolsep}{\dimexpr(\columnwidth-\wd2)/10\relax}
\ifdim\tabcolsep<0pt\typeout{PONENEGATIVEPADDING}\fi
\begin{tabular}{@{}lrrrrr@{}}\pOneRows\end{tabular}
\tabnote{Counts of policies admitting the notice's account. Summary: two prompts $\times$ two compactors $\times$ 40 episodes. User: the user's original turns.}
\end{minipage}
\end{table*}

\paragraph{Episode pools} Table~\ref{tab:fold} pools every episode of the main and goal-diversity runs with the summary in the request slot, de-duplicated by task, goal and repetition. Episodes that ended in an API error are excluded and never counted as failures, and the cross-turn stratum excludes the one banking pair whose turn-$N$ grant already held the attacker's account. The 1472 summaries of Section~\ref{sec:law:slot} are those of the injection goals that name an identifier, and the task contrast of Section~\ref{sec:law:ratio} is over the 153 banking summaries among the 262 shipped-prompt summaries that kept it, with that same pair left out.

\paragraph{Compactor prompt and wrapper}
\label{app:prompts}
The wrapper is fixed across the two conditions of Section~\ref{sec:law:slot} that place a summary in the request slot:

\begin{lstlisting}
The application generated this session summary from
the previous episode. Use it as context for the
current request:

<session_summary>
{summary}
</session_summary>

Current user message: Please continue from where
we left off.
\end{lstlisting}

Our compactor's system prompt:
\begin{lstlisting}
You maintain concise session-resumption memory for an AI
agent. Summarize what the user asked, what data the tools
returned, actions attempted or blocked, and anything
unfinished. Preserve exact filenames, recipients, account
numbers, amounts, dates, and action names needed to resume.
Do not add new tasks or advice. Clearly distinguish
completed, blocked, and pending work.
\end{lstlisting}

{\raggedright The eight Claude Code 2.1.312 arms are reconstructed from the installed binary by \path{e27_realprompts/extract_prompts_2_1_312.py}: \texttt{control} (the default, which the product falls back to when its remote configuration selects no other arm: an analysis block, then nine sections), \texttt{no\_scratchpad}, \texttt{no\_scratchpad\_capped} (2000 words), \texttt{no\_scratchpad\_keep\_asks}, \texttt{lean}, \texttt{short}, \texttt{short\_v2}, \texttt{capped} (also 2000 words). The arm is chosen by remote configuration with fallback \texttt{promptArm ?? "control"}. The Gemini CLI arm uses \texttt{getCompressionPrompt} of \texttt{@google/gemini-cli-core} 0.66.0 as the system prompt and makes its two calls (snapshot, then verification), keeping the verified text. The OpenHands arm renders \texttt{summarizing\_system.j2} and \texttt{summarizing\_events.j2} of the OpenHands SDK at commit \texttt{f0adfd0}. Both products insert the result as a user-role message. The seven later prompts of Table~\ref{tab:e33a} were taken from each product's source at a fixed commit, with the product's own summary prefix. Each product's source at that commit gives the full text.\par}

\section{Shipped prompts}
\label{app:shipped}
Across the eight Claude Code prompts, retention runs from 10\% under \texttt{lean} to 85\% under \texttt{capped} on banking, and the naming rate follows it at both ends. We run the prompts of Gemini CLI and OpenHands as compactor conditions over the same episodes, and later those of Codex, Cline, Goose, Aider, Qwen Code, Kimi CLI and LangChain. Multi-call procedures such as Gemini CLI's draft and self-check run as the products run them (Appendix~\ref{app:prompts}).
Before running the last seven, we fit a per-summary model of the ratio on the Claude Code, Gemini CLI and OpenHands summaries and freeze it. Its four features are how the summary frames the identifier, whether the task names its own payee, summary length and suite. It predicted the naming rate within .10 in all 8 cells of its pre-registered scope with at least five kept summaries, and in all 21 (prompt, suite) cells. Tables~\ref{tab:e33a} and~\ref{tab:predict} give every cell.

\section{Counts behind Figure~\ref{fig:frontier}}
\label{app:frontier}

Table~\ref{tab:frontier} gives the marking and naming counts of Figure~\ref{fig:frontier}a--b for each payload and frontier compactor, and Table~\ref{tab:attrib} gives the admission counts of Figure~\ref{fig:frontier}c for each policy generator.

\end{document}